# A knowledge-based model of civilization under climate change

Boris M. Dolgonosov

Haifa, Israel
E-mail: borismd31@gmail.com

## Abstract

Civilization produces knowledge, which acts as the driving force of its development. A macro-model of civilization that accounts for the effect of knowledge production on population, energy consumption and environmental conditions is developed. The model includes dynamic equations for world population, amount of knowledge circulating in civilization, the share of fossil fuels in total energy consumption, atmospheric $CO_2$ concentration, and global mean surface temperature. Energy dissipation in knowledge production and direct loss of knowledge are taken into account. The model is calibrated using historical data for each variable. About 90 scenarios were calculated. It was shown that there are two control parameters – sensitivity of the population to temperature rise and coefficient of knowledge loss – which determine the future of civilization. In the two-dimensional space of these parameters, there is an area of sustainable development and an area of loss of stability. Calculations show that civilization is located just on the critical curve separating these areas, that is, at the edge of stability. A small deviation can ultimately lead either to a steady state of 10+ billion people or to the complete extinction of civilization. There are no intermediate steady states.

## Highlights

Knowledge is the driving force behind the development of civilization.
Two parameters – temperature sensitivity and knowledge loss – determine civilization development.
In parametric space there exist areas of sustainable development and loss of stability.
Civilization is placed right on the critical curve separating stable and unstable areas.
Civilization eventually comes either to a steady state of 10+ billion or to extinction.

# 1. Introduction

Systematic environmental monitoring reveals global climate change associated with human impact (IPCC, 2013). The increase in global mean surface temperature due to the burning of fossil fuels and some associated processes has numerous negative consequences (Andrews et al., 2018) disrupting the existing human adaptation, which results in deceleration of population growth and may even lead to population decline in the near future. Environmental degradation directs scientific and technological research to reduce fossil fuel consumption, search for new energy sources and develop new environment-friendly technologies. A quantitative description of this system within the framework of a single model is advisable, which however is a difficult problem since it is necessary to combine heterogeneous processes – demographic, physical, technological and informational – and take into account their interaction.

Puliafito et al. (2008) proposed a system of coupled equations describing population dynamics, $CO_2$ emissions, energy consumption and gross domestic product based on an approach in which variables were regarded as separate species interacting as a prey-predator in accordance with the Lotka–Volterra equations. Taagepera (2014) studied the effects of technology and environmental carrying capacity on population growth. Stutz (2014) showed that consumption costs affect the elasticity of the carrying capacity over time. Miranda and Lima (2011) analyzed different approaches to population dynamics and concluded that the power-law model is applicable for the initial stages of the process, while the Allee logistic model can describe the whole process.

The combination of informational and demographic processes within a single model was first undertaken by Dolgonosov and Naidenov (2006), who examined the joint dynamics of knowledge and population. The model is based on the integral principle of least action as applied to the transition of civilization from one level of knowledge to a higher one. The model accounts for the limited resources of the planet, assuming that the environmental conditions are unchanged. However, if we want to describe the human impact on the environment, this assumption must be discarded. Another assumption was that the system accumulates knowledge without loss. But actually, one must keep in mind energy dissipation in knowledge production, and also not neglect the direct loss of knowledge. In order to connect physical processes related to $CO_2$ emissions and temperature rise (Archer et al., 2009; Joos et al. 2013) to this model, we also need to consider the possibility of technological improvements that lower environmental impact by reducing fossil fuel consumption. New technologies are developed using available knowledge that is in active circulation and supported by real knowledge holders. Thus, we come to the concept of active knowledge (introduced by Wiig, 1993), which is used in our study.

A knowledge-based demographic model was considered by Akaev and Sadovnichii (2010), who modified Dolgonosov's (2009) model by including in it delays caused by achieving reproductive age, diffusion of core technologies, and the environmental response to anthropogenic load. Aral (2013, 2014) included an additional term in the modified model that describes the effect of temperature rise on population. The temperature in this model was regarded as an external factor; its change in time was taken from the well-known IPCC scenario allowing a temperature rise of 2°C (Solomon et al., 2007).

Dong et al. (2016) examined the relationship between population $N$ and scientific and technological knowledge $q$ based on an empirical approach. The authors tested several hypotheses regarding this relationship, one of which in our notation looks like this: $\dot{q} = w(q)N$, where $w(q) = aq$ is the per capita knowledge production. Okuducu and Aral (2017) considered five options for the function $w(q)$, including $w(q) =$ constant and a linear dependence $w(q) = aq + b$ with $a = b$. In our approach, we take $w(q) =$ constant, which ensures compliance with the well-known hyperbolic law of population growth empirically discovered by von Foerster et al. (1960).

The objective of this work is to create a model that includes appropriate demographic, physical, technological and informational processes. To reduce complexity, we focus on creating a minimal model that considers dynamics on a global scale, thereby avoiding an increase in the number of parameters while preserving the features of the interaction between the processes.

## 2. Knowledge production and population dynamics

### 2.1. Dissipative dynamics

We have earlier studied the macro-model of civilization as a system that produces the knowledge necessary for survival (Dolgonosov, 2016). Knowledge is defined as a set of descriptions of objects and phenomena (declarative knowledge), as well as a set of process algorithms (procedural knowledge) (ten Berge, van Hezewijk, 1999); for a more detailed classification of knowledge, see Burgin (2017). Strictly speaking, the amount of knowledge is measured by the minimum amount of memory that is needed to store knowledge written in the same language. However, in reality, knowledge is written in different languages without minimizing memory, so we have to deal with the actual amount of memory used.

Based on the least action principle in knowledge production, we have derived the equation of knowledge dynamics

$$\ddot{q} = k\dot{q}^2(1 - b\dot{q}) \qquad (1)$$

where $q$ is the knowledge amount, $k$ is a growth coefficient, and $b$ characterizes the inhibition of knowledge production due to environmental limitations. The inhibition factor $b$ was shown to be a function of knowledge amount

$$b = b_s(1 - e^{-aq}), \qquad (2)$$

where $b_s$ is the inhibition coefficient, and $a$ is the inhibition increment. Equation (1) applies to a system without knowledge loss. However, real civilization not only loses knowledge but also dissipates energy in knowledge production. The dissipation can be taken into account by introducing an additional term $-l\dot{q}$ in equation (1); namely

$$\ddot{q} = k\dot{q}^2(1 - b\dot{q}) - l\dot{q}, \qquad (3)$$

where $l$ is a dissipation factor.

Civilization generates knowledge with the rate

$$\dot{q} = wN, \qquad (4)$$

where $N$ is the population size, and $w$ is the per capita knowledge production rate, which is considered constant as mentioned in Section 1. Substituting (4) in (3), we get a population dynamic equation

$$\dot{N} = kwN^2(1 - bwN) - lN \qquad (5)$$

which depends, through its coefficients, on knowledge amount $q$.

The processing of historical population data shows (Dolgonosov, 2016) that population growth over the last thousand years up to the last quarter of the 20th century is well described by the equation $\dot{N} = kwN^2$ provided that $k$ is constant (same as $w$), which leads to the familiar hyperbolic law (von Foerster et al., 1960).

Based on equations (3) and (4), we can trace the analogy with mechanics, which gives such a correspondence: $q$ is coordinate, $\dot{q}$ is velocity, $N$ is momentum, and $w^{-1}$ is mass. In the spirit of this analogy, the first term on the right in equation (5) is the difference between the driving force $\sim N^2$ (promoting population growth due to knowledge production) and the returning force $\sim N^3$ (inhibiting growth caused by environmental limitations), and the second term is the friction force $\sim N$ (decelerating growth owing to knowledge loss) associated with energy dissipation in a viscous medium.

### 2.2. The concept of active knowledge

The production of knowledge is accompanied by energy dissipation. There are also direct losses of knowledge caused by its obsolescence and knowledge holders' mortality. Only the active knowledge (Wiig, 1993) circulating in society determines the level of life-support technologies and thereby affects the population. This means that the inhibition factor $b$ should depend on the amount of active knowledge $Q$, and not on all the knowledge $q$ produced by civilization in its entire history. Then, instead of equation (2), we must write

$$b = b_s(1 - e^{-aQ}). \tag{6}$$

Knowledge remains active for a limited time. As the population grows, the knowledge lifespan $\tau$ increases, since a proportionately larger number of people can support knowledge. This can be written as

$$\tau = hN, \tag{7}$$

where $h$ is a coefficient.

The balance of active knowledge consists of knowledge production at a rate of $\dot{q} = wN$ and its loss at a rate of $Q/\tau$, which leads to the equation

$$\dot{Q} = wN - \frac{Q}{hN}. \tag{8}$$

A similar equation (in our notation) $\dot{Q} = wN - \lambda Q$, where $Q$ is the amount of culture, was obtained by Ghirlanda et al. (2010), which examined the effect of cultural development on population dynamics. The difference is that in the cited paper $\lambda$ is a constant, whereas according to (8) we have the correspondence $\lambda = 1/(hN)$.

In equilibrium, equation (8) yields $Q = whN^2$, so the active knowledge amount is proportional to the population squared and becomes the larger, the higher the knowledge production rate $w$ and the coefficient $h$. Equation (8) also shows that when civilization collapses and knowledge holders disappear ($N \rightarrow 0$), active knowledge also disappears ($Q \rightarrow 0$). Unlike the active knowledge $Q$, the cumulative amount of knowledge $q$ produced by civilization increases monotonously as long as civilization exists, but if it collapses, $q$ reaches a plateau

$$q_\infty = w\int_0^\infty N(t)dt \qquad (9)$$

(Dolgonosov and Naidenov, 2006) and stays at this level. However, this knowledge is dead because there is no one to utilize it.

## 3. The effect of temperature rise

### 3.1. The dissipation factor

Environmental degradation holds back growth of knowledge production by increasing the dissipation factor $l$ introduced in (3). This parameter depends on environmental conditions through a chain of causal relations: fossil fuel consumption – $CO_2$ emissions – temperature rise – dissipation increase.

On the other hand, the deterioration of environmental conditions induces the chain: knowledge increase – development of technology – reduction in fossil fuel consumption due to the transition to alternative energy sources – decrease in $CO_2$ emissions, temperature and energy dissipation.

Let us consider the temperature dependence of the dissipation factor $l$. As a generalized indicator of environmental conditions, we take global mean surface temperature, although there is an opinion that water scarcity is a critical factor (Parolari et al., 2015). However, most likely it is a secondary factor induced by global warming. As noted by Andrews et al. (2018), heat stress affects the workability and survivability of people, and the dependence of heat exposure on temperature is non-linear. Estimates show that with an increase in temperature by ~2.5℃ above the pre-industrial level about 1 billion people will suffer. The pre-industrial temperature is optimal in the sense that humanity has adapted to it, creating an appropriate life-support infrastructure. Deviations from the optimum will lead to increased energy consumption to maintain a comfortable temperature. Assuming that the dissipation factor is a smooth function of temperature, we can represent it in the vicinity of the optimum as a quadratic form

$$l = l_s \left( 1 + \frac{\Delta T^2}{\sigma^2} \right),$$ (10)

where $l_s$ is the dissipation coefficient, $\Delta T = T - T_0$ is the temperature anomaly, $T$ and $T_0$ are the current temperature and the pre-industrial one, and $\sigma$ may be called "temperature tolerance".

### 3.2. Temperature anomaly

Temperature anomaly is proportional to the radiative forcing, which is a function of atmospheric $CO_2$ concentration. There are several expressions for this function (IPCC, 1990, Chap. 2), the simplest of which is the logarithmic law proposed by Wigley (1987) and leading to

$$\Delta T = \vartheta \ln \frac{C}{C_0},$$ (11)

where $C_0$ and $C$ are the pre-industrial and the current $CO_2$ concentrations, respectively, and $\vartheta$ is a temperature constant. Time series of mean surface temperature and atmospheric $CO_2$ concentration are known from the literature (NASA, 2016; IAC Switzerland, 2014; Scripps UCSD, 2017), which make it possible to find a relationship between temperature and $CO_2$ concentration. This relationship and its trend for 1880 – 2017 are depicted in Fig. 1. The comparison of the logarithmic regression in Fig. 1 with equation (11) gives

$$\vartheta = 3.3917 \text{ K}, \ T_0 - \vartheta \ln C_0 = 267.59.$$ (12)

According to the historical data on carbon dioxide (IAC Switzerland, 2014), $CO_2$ concentration in 1650 is

$$C_0 = 276.41 \text{ ppm}.$$ (13)

Then, from (12) and (13), we get

$$T_0 = 286.66 \text{ K}.$$ (14)

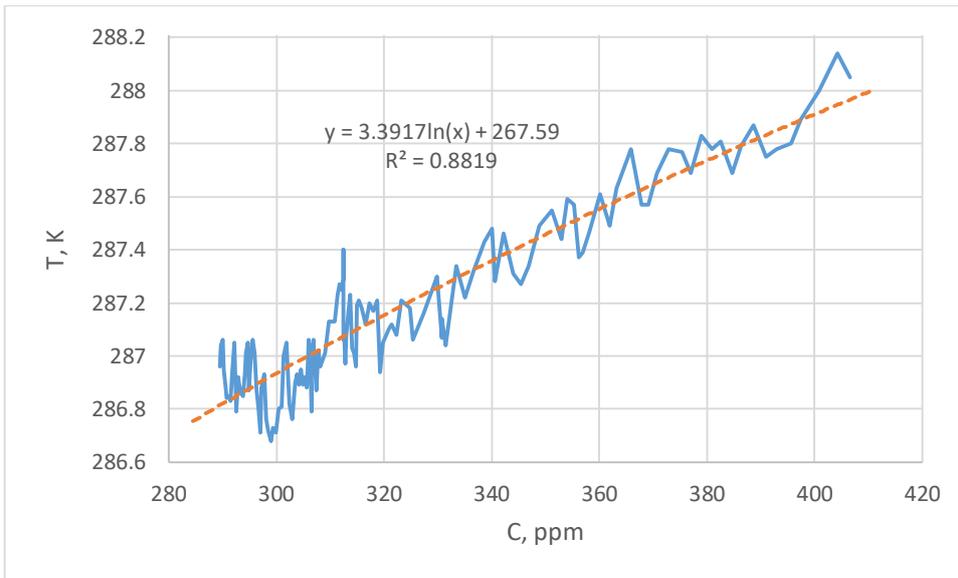

Fig. 1. Earth's mean surface temperature versus $CO_2$ concentration in the atmosphere. Data sources: temperature – NASA, 2016; $CO_2$ – IAC Switzerland, 2014; Scripps UCSD, 2017.

### 3.3. Carbon dioxide emissions

Anthropogenic $CO_2$ emissions superimposed on natural atmospheric $CO_2$ fluctuations lead to an increase in $CO_2$ concentration as shown in Fig. 2 on a logarithmic scale for the period of 1650 – present. A polynomial approximation for $\ln C$ is also indicated there. This approximation is used below in model calibration (Section 5).

Carbon dioxide emissions from burning fossil fuels give an annual increase (ppm/year) in atmospheric $CO_2$

$$J = pE_C, \qquad (15)$$

where $E_C$ is the fossil fuel consumption (Gtoe/year, Gtoe = Giga tonne of oil equivalent), and $p$ is the increase in $CO_2$ per unit of fuel burned (ppm/Gtoe). Coefficient $p$ can be found using the dependence of $CO_2$ concentration on cumulative fuel consumption (Fig. 3). The regression in Fig. 3 indicates that

$$p = 0.2194 \text{ ppm/Gtoe.} \qquad (16)$$

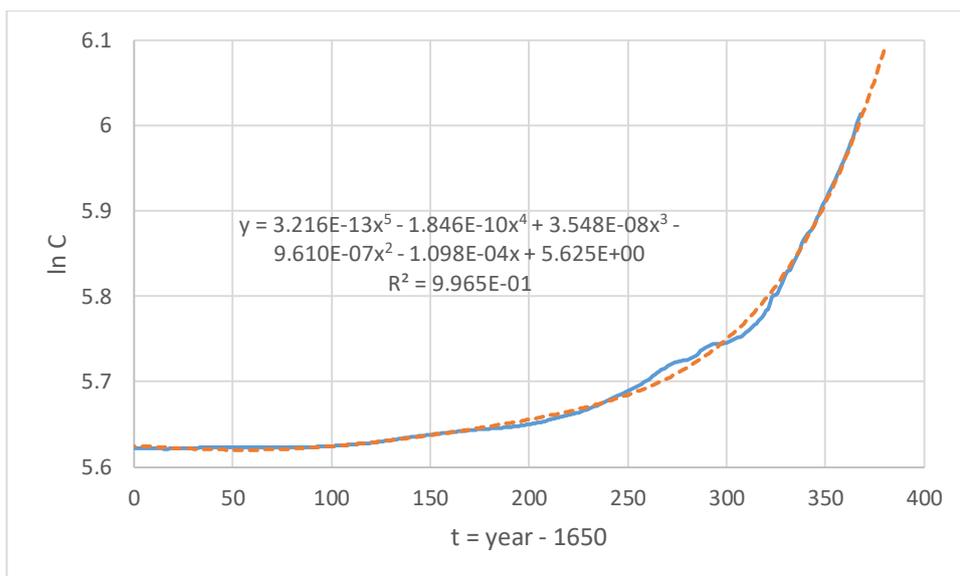

Fig. 2. The logarithm of $CO_2$ concentration (solid curve) and its polynomial approximation (dashed curve). Data sources: IAC Switzerland, 2014; Scripps UCSD, 2017.

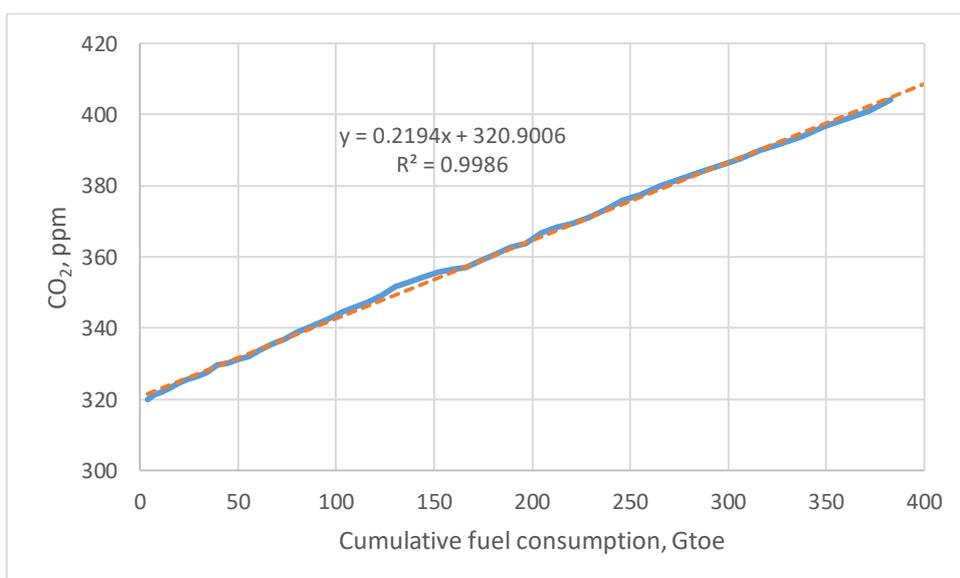

Fig. 3. Atmospheric $CO_2$ concentration versus cumulative fuel consumption. Data sources: $CO_2$ concentration – IAC Switzerland, 2014; energy consumption and $CO_2$ emissions – BP, 2018.

The kinetics of reducing $CO_2$ excess after its instantaneous pulse into the atmosphere was calculated using various models that describe the removal of $CO_2$ through land uptake, ocean invasion and silicate weathering (Archer et al., 2009; Joos et al. 2013). The averaged aggregate kinetics of these processes depicted in Fig. 4 can be approximated by a power-law

$$G(t) = \left(1 + \frac{t}{\tau_C}\right)^{-g}, \quad g = 0.2350, \quad \tau_C = 2.30 \text{ year},\qquad(17)$$

where $G$ is the share of $CO_2$ remaining in the atmosphere, $t$ is the time after the $CO_2$ pulse. For $t \gg \tau_C$, we have the asymptotic law $G(t) \sim t^{-g}$.

Since $CO_2$ emissions occur continuously, $CO_2$ concentration in the atmosphere can be determined by the integral equation

$$C(t) = C_0 + \int_0^t J(\tau)G(t-\tau)d\tau .\qquad(18)$$

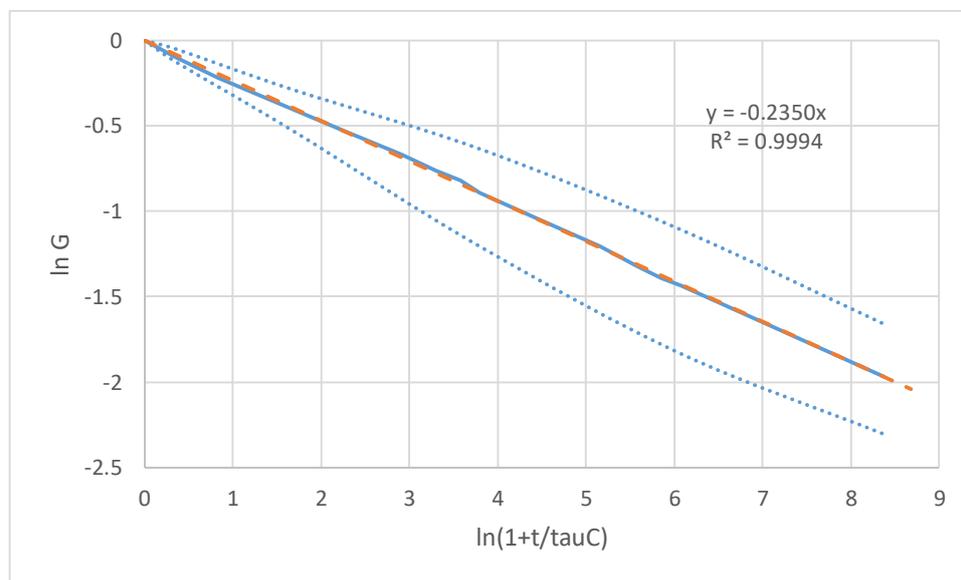

Fig. 4. The share of $CO_2$ pulse remaining in the atmosphere. The initial pulse is 100 GtC at zero moment. Solid curve: kinetics averaged across different models; dotted curves: uncertainty interval ±2 stdev (for calculations on different models); dashed curve: calculations according to equation (17). Data sources: Joos et al., 2013; IPCC, 2013, Chap. 6.

### 3.4. The share of fossil fuels in total energy consumption

The relation between fossil fuel consumption $E_C$ and total energy consumption $E_{tot}$ is

$$E_C = f E_{tot},\qquad(19)$$

where $f$ is the share of fossil fuels in total energy consumption; $f$ depends on the amount of knowledge. As shown by Dolgonosov (2018), total energy consumption is a power-law function of the population (Fig. 5)

$$E_{\text{tot}} = KN^d, \ K = 0.6620, \ d = 1.5, \ R^2 = 0.9886, \tag{20}$$

where $E_{\text{tot}}$ is measured in Gtoe/year and $N$ in billions.

Factor $f$ in equation (19) reflects the human impact on the environment. The accumulation of active knowledge can reduce this impact by phasing out fossil fuels in favor of alternative energy sources. The rate of impact reduction with knowledge growth directly depends on the impact magnitude and the knowledge amount that can be written as

$$f' = -2cQf , \tag{21}$$

where the prime is the derivative with respect to $Q$. A solution to this equation is

$$f = f_0 e^{-cQ^2} . \tag{22}$$

where $f_0$ and $c$ are constants. From this formula it follows that with increasing knowledge, fossil fuel consumption decreases. However, if $Q$ diminishes due to loss of knowledge, fuel consumption resumes its growth.

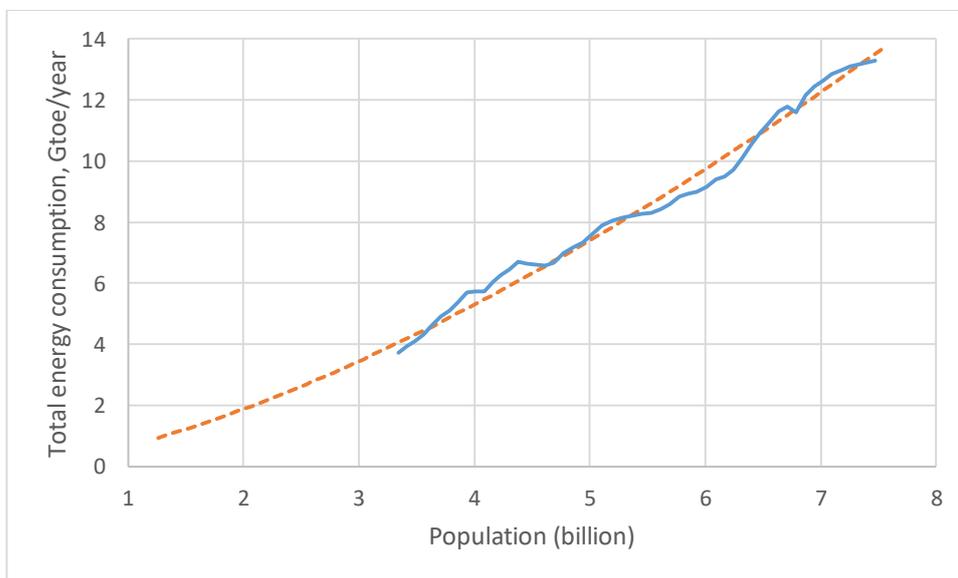

Fig. 5. Total energy consumption as a function of population. Solid curve: data. Dashed curve: trend according to equation (20). Data sources: energy consumption – BP, 2018; population – US Census Bureau, 2018.

## 4. Model

### 4.1. Model equations

Gathering the relationships obtained in the previous sections, we get a complete system of model equations:

active knowledge

$$\dot{Q} = wN - \frac{Q}{hN} ; \tag{23}$$

world population

$$\dot{N} = kwN^2\left(1 - b_s wN(1 - e^{-aQ})\right) - l_s N\left(1 + \frac{\Delta T^2}{\sigma^2}\right) ; \tag{24}$$

temperature anomaly

$$\Delta T = \vartheta \ln \frac{C}{C_0} ; \tag{25}$$

atmospheric CO$_2$ concentration

$$C(t) = C_0 + pf_0 K \int_0^t N^d(\tau) e^{-cQ^2(\tau)} \left(1 + \frac{t - \tau}{\tau_C}\right)^{-g} d\tau . \tag{26}$$

Initial conditions:

$$Q(0) = Q_0, \ N(0) = N_0, \tag{27}$$

where $t = 0$ corresponds to 1650. As shown by Hilbert and Lopez (2012a, 2012b), the amount of information increased by two orders of magnitude over a 21-year interval: from $0.4 \cdot 10^{13}$ Mbyte in 1986 to $30.5 \cdot 10^{13}$ Mbyte in 2007. An even greater difference should be expected when comparing the year 1650 with modernity. The same goes for knowledge, so we take the initial value $Q_0 = 0$.

### 4.2. Normalization

Normalization reduces the number of model parameters. Normalized variables $x, y, z$ and $u$ are defined as

$$t = \tau_s x \,, \ Q = Q_s y \,, \ N = N_s z \,, \ C = C_0(1+u) \,, \tag{28}$$

where the scale factors are

$$\tau_s = \frac{b_s}{k} \,, \ Q_s = \frac{1}{k} \,, \ N_s = \frac{1}{b_s w} \,. \tag{29}$$

Normalized equations:

$$y' = z - \frac{\nu y}{z} \,, \tag{30}$$

$$z' = z^2(1 - z(1 - e^{-\alpha y})) - \gamma z(1 + \beta^2 \ln^2(1+u)) \,, \tag{31}$$

$$u(x) = \delta \int_0^x z^d(s) e^{-\eta y^2(s)} \big(1 + \varepsilon(x-s)\big)^{-g} \, ds \,, \tag{32}$$

where the prime is the derivative with respect to $x$. Initial conditions:

$$y(0) = 0 \,, \ z(0) = z_0 = \frac{N_0}{N_s} \,. \tag{33}$$

Dimensionless parameters in equations (30) – (32) are defined as

$$\alpha = a Q_s \,, \ \beta = \frac{\vartheta}{\sigma} \,, \ \gamma = l_s \tau_s \,, \ \delta = \frac{f_0 p K}{C_0} N_s^d \tau_s \,,$$

$$\eta = c Q_s^2 \,, \ \varepsilon = \frac{\tau_s}{\tau_C} \,, \ \nu = \frac{\tau_s}{h N_s} \,. \tag{34}$$

A brief description of the quantities used in the model (as well as in its calibration, see Section 5) is given in Table 1.

Table 1. Description of the main model variables, parameters, and related quantities

| Notation | Unit | Description |
|---|---|---|
| *Main variables* | | |
| $N$ | billion | world population |
| $C$ | ppm | $CO_2$ concentration in the atmosphere |
| $T$ | K | global mean surface temperature |
| $Q$ | Mbook | active knowledge amount |
| $t$ | year | time |
| $f$ | | share of fossil fuels in total energy consumption |
| | | |
| *Initial values* | | |
| $N_0$ | billion | initial population |
| $C_0$ | ppm | pre-industrial $CO_2$ concentration |
| $T_0$ | K | pre-industrial temperature |
| $Q_0$ | Mbook | initial active knowledge amount |
| $f_0$ | | pre-industrial share of fossil fuels |
| $u_{in}$ | | initial value of relative $CO_2$ excess |
| | | |
| *Dimensionless variables* (see (28) for definitions) | | |
| $x$ | | normalized time |
| $y$ | | normalized active knowledge amount |
| $z$ | | normalized population |
| $u$ | | relative $CO_2$ excess |
| | | |
| *Scale factors* (see (29)) | | |
| $N_s$ | billion | population scale |
| $Q_s$ | Mbook | knowledge amount scale |
| $Q_w$ | Mbook | remainder term in equation (42) |
| $\tau_s$ | year | time scale of population change |
| $\tau_C$ | year | time scale of $CO_2$ concentration change (17) |
| | | |
| *Dimensionless parameters* (see (34)) | | |
| $\alpha$ | | inhibition parameter |
| $\beta$ | | temperature sensitivity |
| $\gamma$ | | dissipation parameter |
| $\delta$ | | $CO_2$ emissions parameter |
| $\varepsilon$ | | population/$CO_2$ timescales ratio |
| $\eta$ | | fuel consumption parameter |
| $\nu$ | | coefficient of knowledge loss |
| | | |
| *Standard deviations in the fitting procedure* | | |
| $s_N$ | billion | s.d. of population |
| $s_f$ | | s.d. of the share of fossil fuels |
| $s_u$ | | s.d. of $CO_2$ excess |
| $s_Q$ | Mbook | s.d. of active knowledge amount |

## 5. Model calibration

The normalized model (30) – (32) includes dimensionless parameters $\alpha$, $\beta$, $\gamma$, $\eta$, $\delta$, $\varepsilon$, $\nu$ (parameters $z_0$, $g$, and $d$ are fixed). To convert them to dimensional ones, we also need to find scale factors $\tau_s$, $Q_s$, and $N_s$. All these parameters are determined by fitting to empirical data, which include the following time series:

- carbon dioxide concentration $C$,
- world population $N$,
- share of fossil fuels $f$ in total energy consumption, and
- book stock $Q_{LC}$ in the Library of Congress as a repository of knowledge.

The calibration procedure is to find parameter values that minimize the deviation of the model from the data. Parameters $\nu$ and $\beta$ are regarded as control ones. Their physical meaning is as follows: $\nu$ is the coefficient of knowledge loss, and $\beta$ is the sensitivity of population to temperature rise. The control parameters are not involved in the calibration and can freely vary in certain ranges. The other parameters must satisfy the minimum deviation condition, which turns them into functions of $\nu$ and $\beta$. To reduce uncertainty, the calibration is divided into four steps.

### 5.1. Step 1: Population

The subset $P = \{\alpha, \gamma, N_s, \tau_s\}$ of model parameters is considered. The minimum variance is determined as

$$V_N(\nu, \beta) = \min_P \sum_{i=1}^{n} \omega_i \left( N(t_i; P, \nu, \beta) - N_{\text{aprx}}(t_i) \right)^2, \tag{35}$$

where

$$N(t) = N_s z(x), \ \ x = \frac{t}{\tau_s}, \ \ \omega_i = \frac{t_i - t_{i-1}}{t_n - t_0}, \tag{36}$$

$\omega_i$ is the weight factor, $n$ is the number of points. Function $z(x)$ is found by solving a pair of equations (30) and (31) with initial conditions

$$y(0) = 0 , \; z(0) = z_0 = 0.055. \tag{37}$$

Function $u(x)$ presented in (31) is calculated here as $u = C_{aprx}/C_0 - 1$ (instead of (32)), where $C_{aprx}$ is defined by the two regressions shown in Fig. 2. In equation (35), $N_{aprx}(t)$ is the population time series approximation indicated in Fig. 6. At the output of this step, we obtain the optimal parameters' values: $P = P_{opt}(\nu, \beta)$.

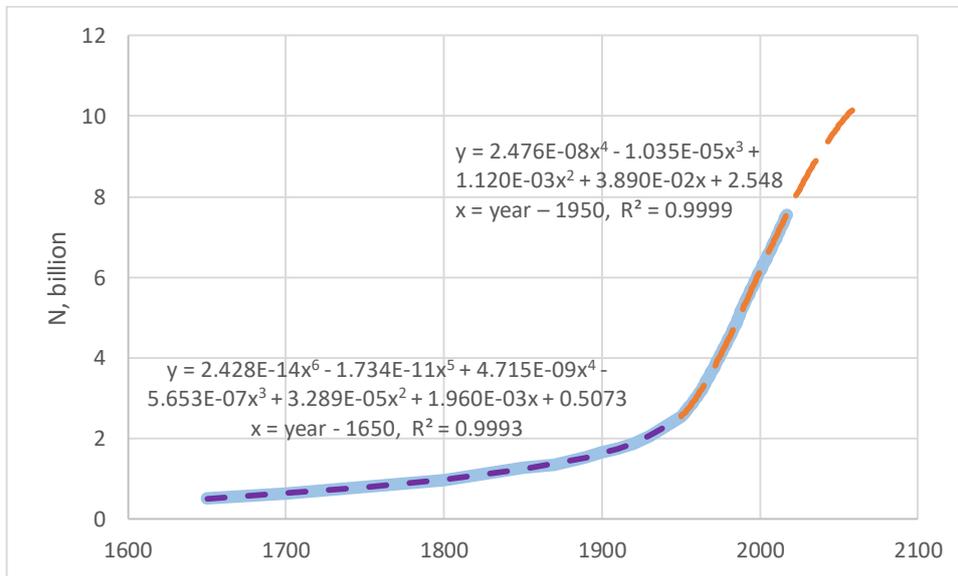

Fig. 6. Population time series (solid curve) and trends in the intervals of 1650 – 1950 and 1950 – 2017 (dashed curves). Data source: US Census Bureau, 2018.

## 5.2. Step 2: Share of fossil fuels

Now we consider the subset of parameters $P = \{f_0, \eta\}$ belonging to the function

$f(x) = f_0 e^{-\eta y^2(x)}$ (see Section 3.4). The minimum variance is

$$V_f(\nu, \beta) = \min_P \sum_{i=1}^{n} \omega_i \left( f(x_i; P, \nu, \beta) - f_{aprx}(t_i) \right)^2 , \tag{38}$$

where $x_i = t_i / \tau_s$; $f_{\mathrm{aprx}}(t)$ is the data approximation specified in Fig. 7. Function $y(x)$ present in the expression for $f(x)$ is found from equation (30), in which $z$ is calculated as $z(x) = N_{\mathrm{aprx}}(t) / N_s$ instead of equation (31). As before, the optimal parameters' values depend on $\nu$ and $\beta$.

Approximation in Fig. 7 represents the envelope, which is built according to the data for 1960–1973, 2012–2016, and 2040 (the last point is BP projection for 2040). Data for 1974 – 2011 correspond to the growth of nuclear energetics (Fig. 8), which by 2011 had ceased because of safety problems. These data are not used for calibration since they describe a temporary deviation from the trend; this deviation is associated with unjustified expectations regarding nuclear power. Over time, the share of fossil fuels returned to the level provided by relatively safe energy production technologies. Along with the BP projection until 2040, our projection for 2050, obtained by extrapolating the trends of various energy sources, is presented in Fig. 7; the trends are shown in Fig. 8 and described in Table 2. It can be seen that our projection is consistent with the BP one.

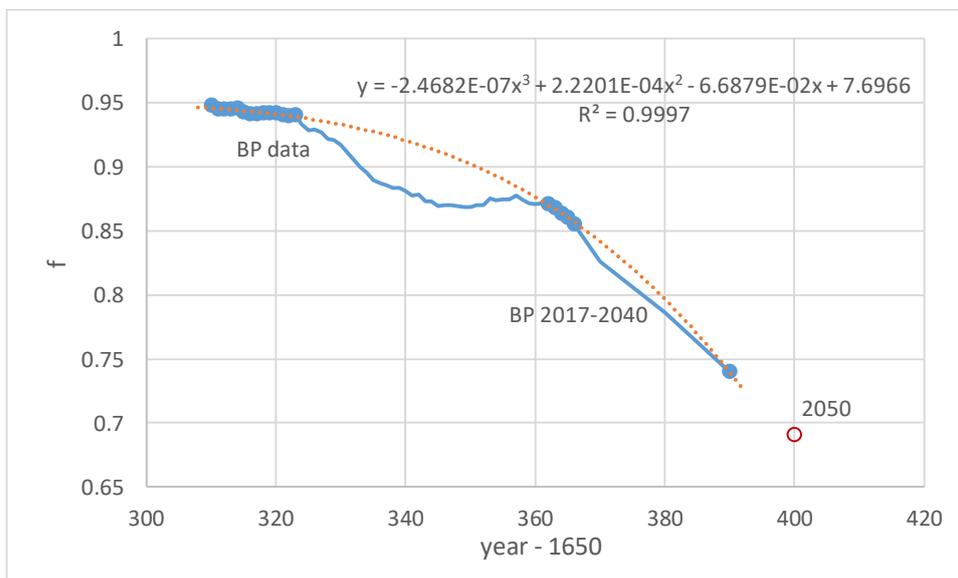

Fig. 7. The share of fossil fuels in total energy consumption and its trend (dotted curve). Filled circles: data used to find the regression. Empty circle: our projection for 2050. Data source: BP, 2018.

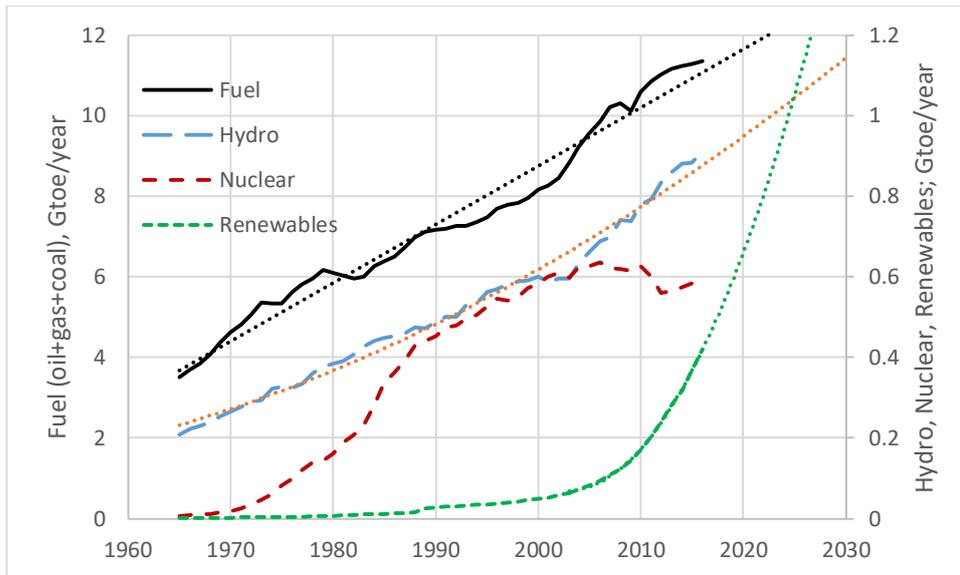

Fig. 8. Time series of various sources of energy consumption and their trends (dotted curves). The trends are described in Table 2. Data source: BP, 2018.

Table 2. World energy consumption (Gtoe/year) in 2016 and projection for 2050

| Energy sources | 2016** | % total | Trends (x=year, y=energy consumption) | 2050 | % total |
|---|---|---|---|---|---|
| Fuel (oil+gas+coal) | 11.35 | 85.5 | y = 0.1448x - 280.89, $R^2$ = 0.9721 | 15.97 | 69.1 |
| Hydro | 0.91 | 6.9 | y = 9.942E-05$x^2$ - 3.832E-01x + 3.693E+02, $R^2$ = 0.9863 | 1.51 | 6.6 |
| Nuclear | 0.59 | 4.4 | Estimate | 1.00 | 4.3 |
| Renewables* | 0.42 | 3.2 | y = 1.982E-03$x^2$ - 7.937E+00x + 7.948E+03, $R^2$ = 0.9995 | 4.62 | 20.0 |
| **Total** | **13.27** | **100** | | **23.10** | **100** |

*Gross generation from renewable sources includes wind, geothermal, solar, biomass and waste.
**Data source: BP, 2018.

## 5.3. Step 3: $CO_2$ concentration

In this step, we consider the subset of parameters $P = \{\varepsilon, u_{in}\}$. Calibration is carried out in the interval 1800 – present, because earlier data are rare and unreliable. Since the initial moment has shifted from 1650 to 1800, it is necessary to change the $CO_2$ excess initial value $u(0)$. In (32) there was $u(0) = 0$, and now it should be $u(0) = u_{in}$; the term $u_{in}$ must be added to equation (32) in front of the integral. The minimum variance is

$$V_C(\nu, \beta) = \min_P \sum_{i=1}^{n} \omega_i \Big( C(t_i; P, \nu, \beta) - C_{\mathrm{aprx}}(t_i) \Big)^2 , \qquad (39)$$

where $C(t) = C_0(1 + u(x))$, $x = t/\tau_s$; $C_{\mathrm{aprx}}(t)$ is calculated using the regression shown in Fig. 1. Function $u(x)$ is found by solving a pair of equations (30) and (32). In (30), variable $z$ is calculated as $z(x) = N_{\mathrm{aprx}}(t)/N_s$, where $x = t/\tau_s$; $N_{\mathrm{aprx}}(t)$ is shown in Fig. 6. Equation (30) is solved in the interval 1650 – present with the initial condition $y(0) = 0$ corresponding to 1650.

The aforementioned shift of the initial moment from 1650 to 1800 in the calculation of $C$ while all other model equations are calculated starting from 1650, leads to the following change: parameter $\tau_C$ participating in the definition of $\varepsilon = t_s / \tau_C$ takes on a different value than that specified in (17). The new value of $\tau_C$ must be of the same order as the temporal shift ($\sim 10^2$ years). However, the exponent $g$ in the power-law (17) retains its value, leaving the asymptotic behavior unchanged.

### 5.4. Step 4: Knowledge

Now we can calculate dynamics of the variables $x, y, z$ using equations (30) – (32). The relationship between $y$ and $Q$ is

$$Q(t) = Q_s y(x) , \quad x = \frac{t}{\tau_s} . \qquad (40)$$

Dynamics of knowledge accumulation is reflected in the development of the largest repository of knowledge – the Library of Congress (LC) (Fig. 9). To calibrate the model, we used data on the size of the book stock (number of volumes) for 1898 – 1982. During this period, the report form did not change. The change happened after 1982, when an increasing part of the Library's budget begins to be spent on computer carriers, so the method of assessing knowledge growth only by books becomes inadequate since it underestimates the actual increase in knowledge.

We assume a linear relationship between world knowledge amount $Q$ and the LC book stock $Q_{\mathrm{LC}}$:

$$Q(t) = \kappa Q_{\mathrm{LC}}(t) + Q_w , \qquad (41)$$

where $\kappa$ is a coefficient converting the number of books in LC into the total number of books in the world, and $Q_w$ is a remainder term. The LC book stock records are approximated by polynomial regression, shown in Fig. 9. Since there are no data to calculate the conversion coefficient $\kappa$ in (41), we put $\kappa = 1$, thereby measuring the amount of knowledge in the "LC book equivalent" units (further for short we write simply "book"). In these units we have

$$Q(t) = Q_{LC}(t) + Q_w. \tag{42}$$

The minimum variance is

$$V_Q(\nu, \beta) = \min_{Q_s, Q_w} \sum_{i=1}^{n} \omega_i \left( Q_s\, y(x_i; \nu, \beta) - Q_{LC}(t_i) - Q_w \right)^2. \tag{43}$$

The optimum of $Q_s$ and $Q_w$ is a function of $\nu$ and $\beta$.

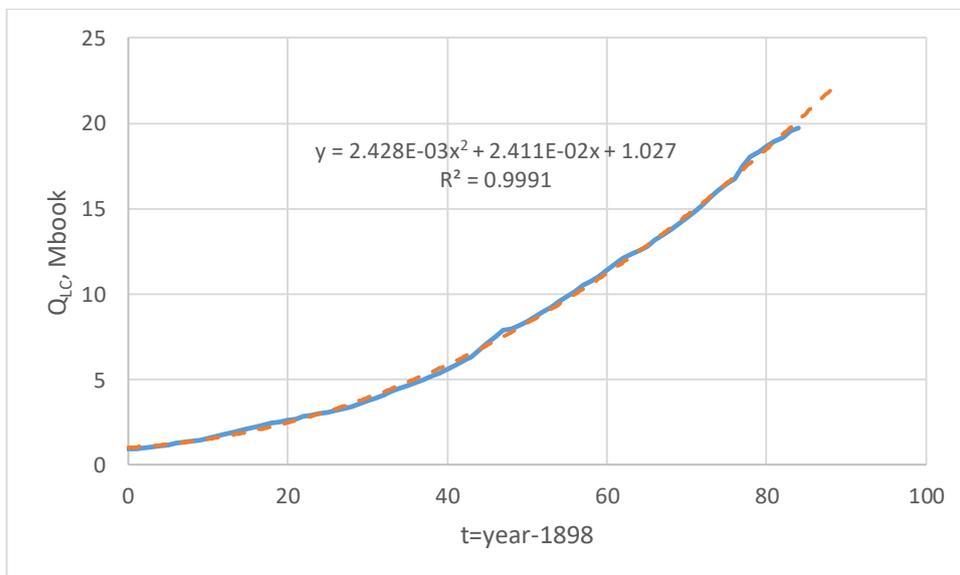

Fig. 9. Book stock of the Library of Congress for 1898 – 1982 (solid curve: data; dashed curve: trend). Only books (volumes, excluding pamphlets) are counted. The three-year gap between 1942 and 1944 was filled by linear interpolation. Data source: Library of Congress, 2017.

## 6. Results

The four-step procedure of minimizing deviations determines the dimensionless parameters $\alpha$, $\gamma$, $\delta$, $\varepsilon$, $\eta$, $f_0$, scale factors $N_s$, $\tau_s$, $Q_s$, and parameter $Q_w$, each as a function of $\nu$ and $\beta$.

The calculations according to equations (30) – (32) were carried out using the following constant parameters collected from the previous sections:

$$\vartheta = 3.3917 \text{ K}, \ C_0 = 276.41 \text{ ppm}, \ T_0 = 286.66 \text{ K},$$

$$p = 0.2194 \text{ ppm/Gtoe}, \ K = 0.6620 \text{ Gtoe/year/billion}^{1.5}, \qquad (44)$$

with initial conditions $y_0 = 0$, $z_0 = 0.055$ for $t = 0$ (year 1650).

The dynamics were calculated for the following variables: population $N$, relative $CO_2$ excess $u = C / C_0 - 1$, temperature anomaly $\Delta T$, and active knowledge amount $Q$. About 90 different scenarios were calculated to a depth of several thousand years, and if necessary (e.g., to determine the critical curve; see below) up to 1 million years.

## 6.1. The critical curve

All scenarios end with either a finite steady state or the collapse of civilization with its complete disappearance. The critical curve $\beta(\nu)$ separating these two final states is shown in Fig. 10. Its intersections with the axes are $\beta(0) = \beta^* = 3.415$ and $\beta(\nu^*) = 0$, where $\nu^* = 0.0467$ is the endpoint beyond which only such a civilization can exist that is insensitive to temperature rise.

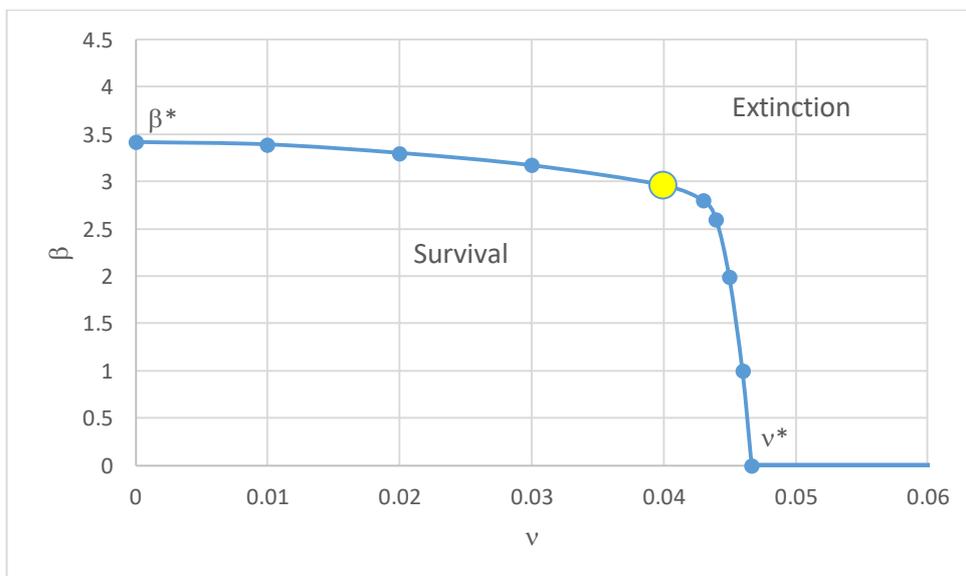

Fig. 10. The critical curve $\beta(\nu)$, separating the areas where civilization survives or perishes. The highlighted point shows the most probable place of civilization (see Section 7).

In the case of zero loss of knowledge ($\nu = 0$), the existence of civilization is possible in the widest range of $\beta$: from absolute temperature insensitivity $\beta = 0$ to high sensitivity $\beta = \beta^*$. With increasing knowledge loss, the admissible sensitivity range gradually narrows, and after passing $\nu = 0.043$, the sensitivity on the critical curve rapidly decreases and vanishes at $\nu = \nu^*$.

Figure 11 shows the lifetime of civilization in the supercritical area. Four cross-sections of the critical curve corresponding to the loss coefficient $\nu = 0$, 0.04, 0.05 and 0.1 are presented. At high sensitivity, lifetime is short: for $\beta = 4$, extinction occurs between 2300 and 2400. As sensitivity decreases, lifetime increases, tending to infinity when approaching the critical curve. For example, collapse occurs after 280 thousand years for $\nu = 0.05$ and $\beta = 2.5$, and after 38 thousand years for $\nu = 0.1$ and $\beta = 1$ (both lifetimes are far beyond the graph). Smaller $\beta$ values correspond to lifetimes in millions of years. In this case, we can regard civilization as metastable since it is in the supercritical area and is doomed to extinction, albeit in a very distant future (however, there is hope that for such a long time, favorable changes can occur in civilization or the environment, but this matter is beyond the scope of the model).

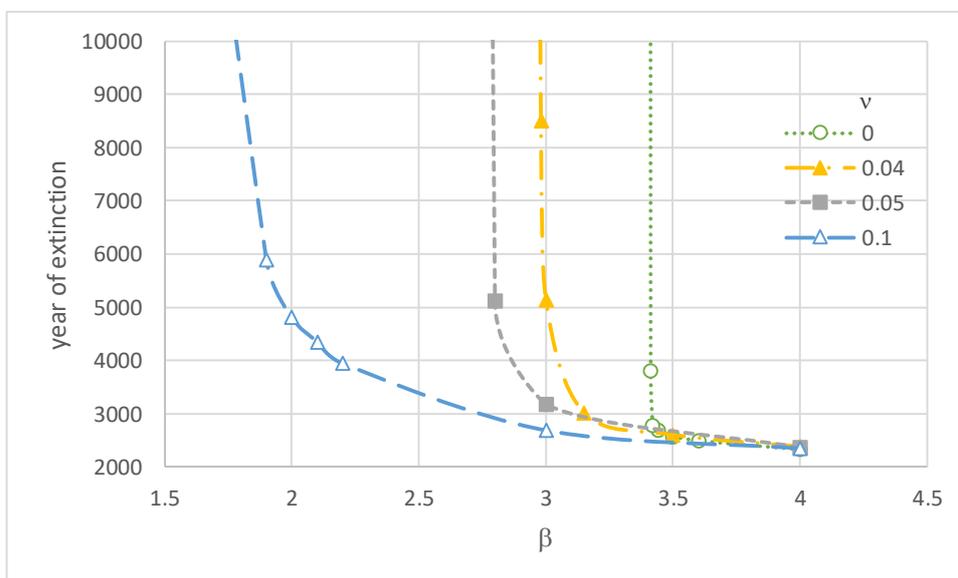

Fig. 11. The year of extinction, defined as the year when the population falls below 0.1 billion.

*6.2. Development scenarios*

Let us consider several calculated scenarios of the transition through the critical curve in three cross-sections $\nu = 0.01$, 0.04, and 0.1 (Figs. 12 – 14). Each scenario is characterized by two parameters: knowledge loss coefficient $\nu$ and sensitivity of population to temperature rise $\beta$. Information on scenarios with different values of $\beta$ in these cross-sections is presented in Table 3.

*6.2.1. Low knowledge loss*

Scenarios with a low loss coefficient $\nu = 0.01$ are presented in Fig. 12. The left panels show good agreement with empirical data for all scenarios, regardless of temperature sensitivity. The scenarios diverge over time. This is especially noticeable concerning the population, which in different scenarios varies significantly by the mid-century. In a more distant future, the scenarios radically diverge as shown in the right panels which demonstrate dynamics up to 5600.

In the case of absolute insensitivity to temperature rise ($\beta = 0$), the population by the end-century almost reaches a plateau of 10.1 billion people. A little later, $CO_2$ excess and temperature reach their maximums $u = 1.34$ and $T = 16.3°C$; the latter is 1.4°C higher than the temperature of 2017 (14.9°C). At the same time, active knowledge amount is growing rapidly.

With increasing temperature sensitivity, population dynamics become more complex. For $\beta = 2.8$, the population reaches a maximum of 9.3 billion in 2058, then over two centuries, population drops to a minimum of 8.1 billion in 2251. After that, growth resumes, asymptotically tending to a plateau of 10.3 billion.

A similar situation holds for the pre-critical sensitivity $\beta = 3.39$. The population maximum of 9 billion falls in 2050, and the minimum becomes deeper, 4.8 billion in 2512. Then there is a slow growth to a plateau of 10.4 billion.

A barely noticeable increase in sensitivity $\beta$ from 3.39 to 3.40 leads to catastrophic consequences: after a maximum of 9 billion in 2050, there is a decline in the population, which until 2300 practically coincides with the previous non-catastrophic scenario. However, further the decline does not stop, and by 2940 the population falls below the level of 0.1 billion (which we conditionally accept as the moment of the disappearance of civilization).

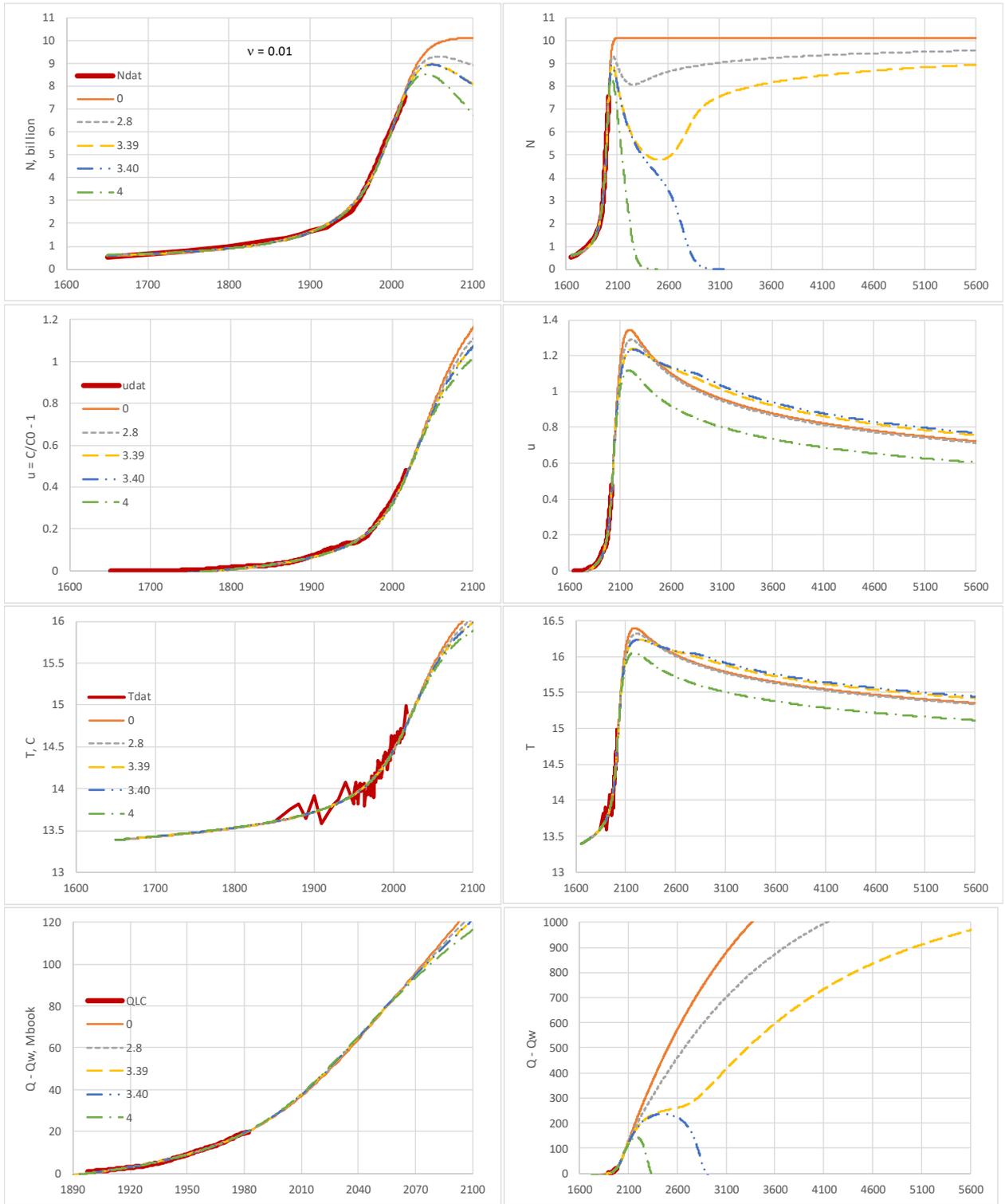

Fig. 12. Dynamics of population, CO₂ excess, temperature, and active knowledge amount: scenarios with loss coefficient ν = 0.01. Legend: data and temperature sensitivity β values. Left panels: projections until 2100 and comparison with historical data. Right panels: projections until 5600.

Even more rapid disappearance of civilization occurs for high sensitivity $\beta = 4$: a maximum of 8.6 billion is reached in 2042, and the collapse occurs as early as 2348.

In the above scenarios, the temperature reaches a maximum of $16 - 16.5°C$ around 2200, and then it slowly decreases due to a decrease in $CO_2$ excess over tens of thousands of years (tails relax even longer, over hundreds of thousands of years).

The amount of active knowledge (more precisely, the difference $Q - Q_w$) in subcritical scenarios ($\beta = 0$, 2.8, and 3.39) increases reaching a high plateau ($\sim 1500$ Mbook, Table 3), whereas in supercritical scenarios ($\beta = 3.4$ and 4), active knowledge goes through a maximum (an order of magnitude lower than the mentioned plateau), and then disappears simultaneously with the collapse of civilization.

### 6.2.2. Medium knowledge loss

As the loss coefficient increases, scenarios change. The situation for $\nu = 0.04$ is shown in Fig. 13. The left panels show good agreement with the data. The discrepancy between different scenarios become noticeable since the mid-century. The transition through the critical curve occurs for $\beta$ between 2.96 and 2.97. The right panels show a dramatic change in scenarios when crossing the critical curve. In the subcritical area, after passing through a maximum and a minimum, the population increases reaching a plateau (slightly above 10 billion, scenarios $\beta = 2.7$, 2.9, 2.96). In the supercritical area (scenarios $\beta > 2.96$), population eventually collapses; its lifetime is very long near the critical curve and decreases with distance from it (with increasing $\beta$) as shown in Fig. 11 and Table 3. The scenario $\beta = 2.97$ is not shown in Fig. 13, since in the time interval depicted it is practically indistinguishable from the scenario $\beta = 2.96$, but deviates strongly from the latter at a more distant time, experiencing a collapse after almost 20 thousand years.

The temperature in the subcritical area ($\beta \leq 2.96$) reaches a maximum (about $16.3°C$), and then slowly decreases due to a decrease in $CO_2$ excess. In the supercritical area (scenario $\beta > 2.96$), the temperature behavior depends on the proximity to the critical curve. In its vicinity, temperature increases (possibly after a certain decrease, as for $\beta = 2.97$ and 3) and eventually becomes intolerable for the population, which begins to decline in size rapidly, leading to a decrease in $CO_2$ emissions and temperature. The second temperature maximum for $\beta = 2.97$ (equal to 17.1℃) is not visible in Fig. 13, since it falls on the year 20181, after which civilization quickly collapses (in 20292). For $\beta = 3$, the second temperature maximum (also about 17.1℃) is much closer, namely in 5007, and the collapse year is 5127. With very high sensitivity (scenario $\beta = 4$), the population drops so quickly that distant temperature peaks disappear, and the temperature begins to decline immediately after passing the first maximum of 16.2℃ in 2249, and after 129 years a collapse occurs. Note that $CO_2$ excess has the same dynamics.

Active knowledge in the subcritical area grows, reaching a plateau (568, 495, and 463 Mbook for the first three scenarios in Fig. 13), which is significantly lower than that in the corresponding group of scenarios for $\nu = 0.01$. In the supercritical area, active knowledge is nullified along with the disappearance of civilization.

*6.2.3. High knowledge loss*

With a high value of the loss coefficient $\nu = 0.1$, the only scenario that preserves civilization corresponds to absolute insensitivity to temperature rise, $\beta = 0$. Any finite sensitivity leads to the collapse of civilization over time. This situation is illustrated in Fig. 14, which presents several scenarios with different $\beta$ values. For $\beta > 0$ and a high loss of knowledge, civilization is not able to create environment-friendly life-support technologies that would ensure sustainable development. Under increasing heat stress, the population is declining. The comparative population and $CO_2$ dynamics (right panels $N$ and $u$ in Fig. 14) show that shortly after passing the $CO_2$ maximum (which coincides with the temperature one), civilization collapses. In parallel with population decline, the active knowledge amount is also decreasing. Details are presented in Table 3.

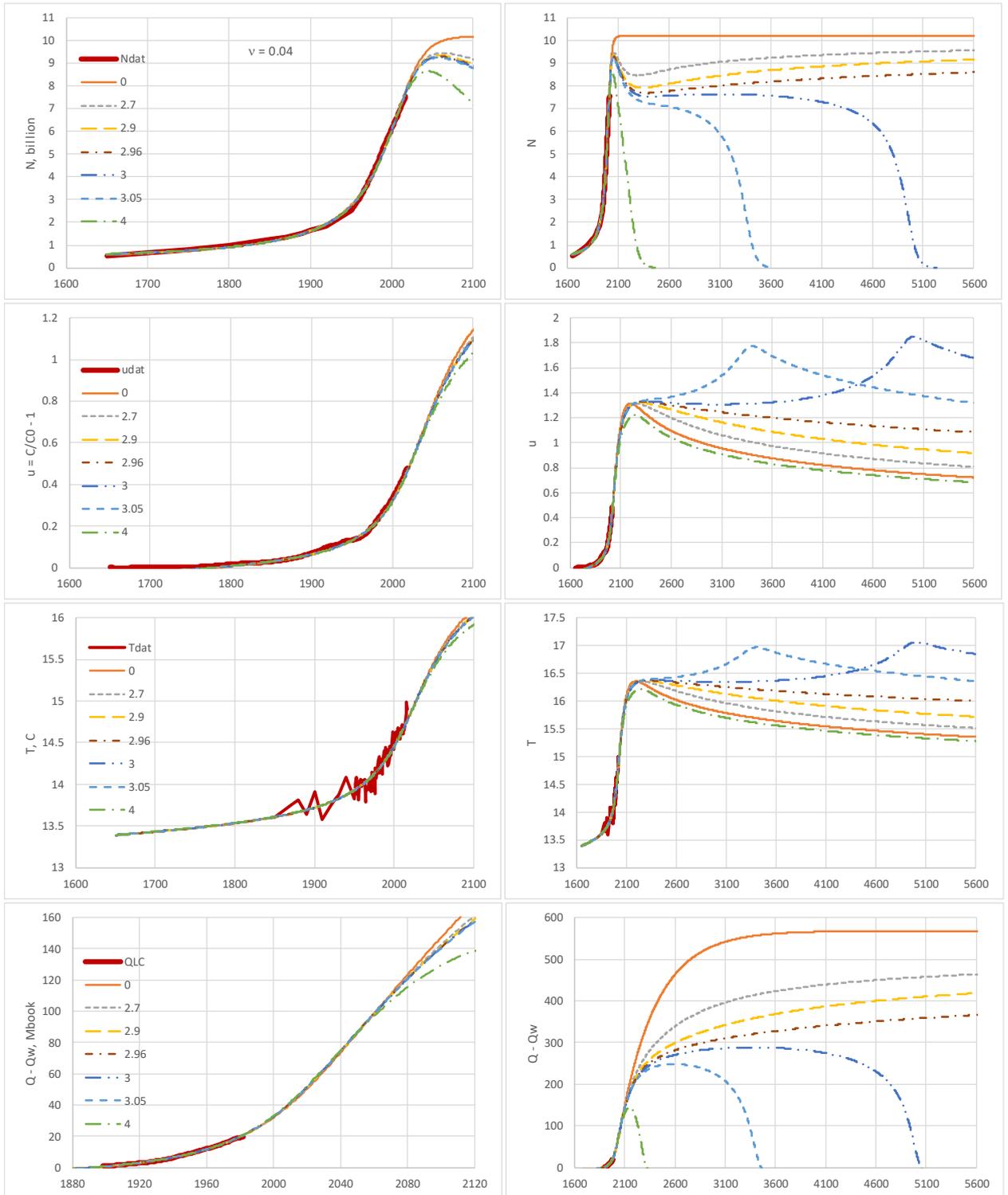

Fig. 13. Scenarios with loss coefficient $\nu = 0.04$.

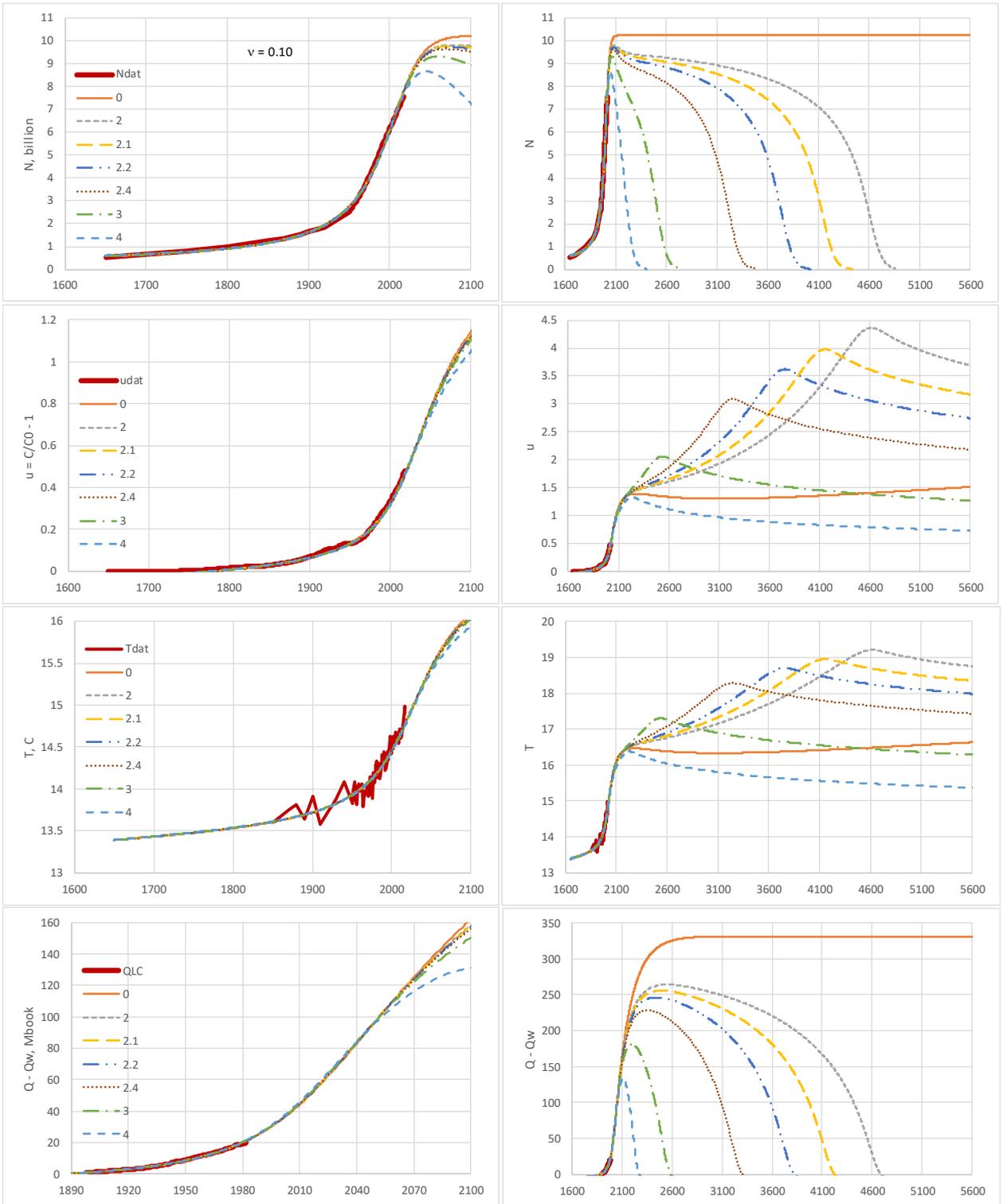

Fig. 14. Scenarios with loss coefficient ν = 0.1.

Table 3. Characteristics of selected scenarios*

| ν | β | α | γ | N_s | τ_s | Q_s | Q_w | f_0 | η | τ_C | u_in | δ | ε | N maximum** | | N minimum | | N plateau | Extinction | T maximum** | | (Q-Q_w) maximum‡ | |
|---|---|---|---|---|---|---|---|---|---|---|---|---|---|---|---|---|---|---|---|---|---|---|---|
| | | | | billion | year | Mbook | Mbook | | | year | | | | year AD | billion | year AD | billion | billion | year AD | year AD | °C | year AD | Mbook |
| 0.01 | 4 | 1.08 | 0.0345 | 11.2 | 11.5 | 16.0 | 13.19 | 0.977 | 0.0105 | 108 | -0.0355 | 0.220 | 0.106 | 2042 | 8.6 | | | 0 | 2348 | 2187 | 16.1 | 2182 | 145 |
| | 3.40 | 1.22 | 0.0319 | 10.9 | 12.1 | 16.0 | 13.44 | 0.978 | 0.0109 | 107 | -0.0355 | 0.225 | 0.113 | 2050 | 9.0 | | | 0 | 2940 | 2228 | 16.2 | 2457 | 237 |
| | 3.39 | 1.20 | 0.0317 | 10.9 | 12.2 | 16.1 | 13.47 | 0.978 | 0.0110 | 107 | -0.0356 | 0.226 | 0.113 | 2050 | 9.0 | 2512 | 4.8 | 10.4 | | 2228 | 16.2 | | 1447 |
| | 2.8 | 1.26 | 0.0294 | 10.7 | 12.7 | 16.2 | 13.69 | 0.979 | 0.0114 | 107 | -0.0356 | 0.231 | 0.119 | 2058 | 9.3 | 2251 | 8.1 | 10.3 | | 2212 | 16.3 | | 1484 |
| | 0 | 1.26 | 0.0242 | 10.4 | 14.0 | 16.7 | 14.14 | 0.981 | 0.0125 | 106 | -0.0356 | 0.242 | 0.132 | | | | | 10.1 | | 2191 | 16.4 | | 1573 |
| 0.04 | 4 | 1.44 | 0.0316 | 11.1 | 12.6 | 24.1 | 7.55 | 0.959 | 0.0211 | 114 | -0.0343 | 0.234 | 0.110 | 2044 | 8.7 | | | 0 | 2378 | 2249 | 16.2 | 2149 | 144 |
| | 3.75 | 1.56 | 0.0317 | 11.1 | 12.5 | 23.9 | 7.53 | 0.959 | 0.0209 | 114 | -0.0344 | 0.233 | 0.109 | 2047 | 8.8 | | | 0 | 2427 | 2293 | 16.3 | 2167 | 155 |
| | 3.6 | 1.59 | 0.0311 | 11.0 | 12.7 | 23.9 | 7.59 | 0.959 | 0.0211 | 114 | -0.0344 | 0.234 | 0.111 | 2050 | 8.9 | | | 0 | 2484 | 2346 | 16.4 | 2187 | 165 |
| | 3.45 | 1.61 | 0.0305 | 11.0 | 12.8 | 23.9 | 7.65 | 0.959 | 0.0212 | 114 | -0.0344 | 0.235 | 0.112 | 2051 | 9.0 | | | 0 | 2566 | 2426 | 16.5 | 2215 | 178 |
| | 3.3 | 1.62 | 0.0298 | 10.9 | 13.0 | 23.9 | 7.72 | 0.959 | 0.0214 | 114 | -0.0344 | 0.237 | 0.114 | 2054 | 9.1 | | | 0 | 2707 | 2567 | 16.6 | 2262 | 195 |
| | 3.15 | 1.64 | 0.0292 | 10.9 | 13.1 | 23.9 | 7.77 | 0.959 | 0.0215 | 114 | -0.0344 | 0.238 | 0.115 | 2056 | 9.2 | | | 0 | 2990 | 2854 | 16.8 | 2359 | 219 |
| | 3.05 | 1.65 | 0.0288 | 10.9 | 13.2 | 23.9 | 7.81 | 0.960 | 0.0216 | 114 | -0.0345 | 0.239 | 0.116 | 2057 | 9.3 | | | 0 | 3535 | 3406 | 17.0 | 2594 | 248 |
| | 3 | 1.64 | 0.0285 | 10.8 | 13.3 | 23.9 | 7.84 | 0.960 | 0.0217 | 113 | -0.0344 | 0.239 | 0.117 | 2058 | 9.3 | 2427 | 7.5 | 0 | 5127 | 2354 | 16.4 | 3376 | 288 |
| | 2.97 | 1.65 | 0.0284 | 10.8 | 13.3 | 23.9 | 7.85 | 0.960 | 0.0217 | 114 | -0.0345 | 0.239 | 0.117 | 2058 | 9.3 | 2391 | 7.7 | 0 | 20292 | 2341 | 16.4 | 11174 | 363 |
| | 2.96 | 1.65 | 0.0284 | 10.8 | 13.3 | 23.9 | 7.85 | 0.960 | 0.0217 | 114 | -0.0345 | 0.239 | 0.117 | 2058 | 9.3 | 2381 | 7.7 | 10.2 | | 2320 | 16.4 | | 516 |
| | 2.9 | 1.65 | 0.0281 | 10.8 | 13.4 | 23.9 | 7.88 | 0.960 | 0.0218 | 114 | -0.0345 | 0.240 | 0.118 | 2060 | 9.3 | 2339 | 7.9 | 10.3 | | 2289 | 16.4 | | 536 |
| | 2.7 | 1.65 | 0.0273 | 10.8 | 13.6 | 23.9 | 7.95 | 0.960 | 0.0219 | 114 | -0.0345 | 0.242 | 0.119 | 2062 | 9.5 | 2288 | 8.5 | 10.3 | | 2251 | 16.4 | | 544 |
| | 1.5 | 1.65 | 0.0240 | 10.5 | 14.4 | 24.1 | 8.25 | 0.961 | 0.0228 | 113 | -0.0344 | 0.249 | 0.128 | 2086 | 10.0 | 2225 | 9.8 | 10.2 | | 2203 | 16.4 | | 550 |
| | 0.7 | 1.64 | 0.0228 | 10.5 | 14.7 | 24.1 | 8.36 | 0.961 | 0.0230 | 113 | -0.0345 | 0.251 | 0.130 | 2117 | 10.14 | 2216 | 10.13 | 10.2 | | 2197 | 16.4 | | 564 |
| | 0 | 1.63 | 0.0225 | 10.4 | 14.8 | 24.2 | 8.39 | 0.961 | 0.0231 | 113 | -0.0345 | 0.252 | 0.131 | | | | | 10.2 | | 2196 | 16.4 | | 569 |
| 0.1 | 4 | 2.10 | 0.0313 | 11.1 | 12.8 | 36.0 | 4.89 | 0.951 | 0.0393 | 118 | -0.0339 | 0.238 | 0.108 | 2046 | 8.7 | | | 0 | 2349 | 2235 | 16.4 | 2111 | 133 |
| | 3 | 2.27 | 0.0273 | 10.8 | 13.8 | 35.5 | 5.18 | 0.951 | 0.0399 | 118 | -0.0340 | 0.246 | 0.117 | 2060 | 9.3 | | | 0 | 2689 | 2536 | 17.3 | 2197 | 181 |
| | 2.4 | 2.27 | 0.0251 | 10.7 | 14.3 | 35.3 | 5.34 | 0.952 | 0.0404 | 118 | -0.0340 | 0.251 | 0.122 | 2070 | 9.6 | | | 0 | 3421 | 3242 | 18.3 | 2352 | 229 |
| | 2.2 | 2.28 | 0.0246 | 10.7 | 14.5 | 35.3 | 5.37 | 0.952 | 0.0405 | 117 | -0.0339 | 0.252 | 0.123 | 2074 | 9.7 | | | 0 | 3946 | 3756 | 18.7 | 2439 | 246 |
| | 2.1 | 2.27 | 0.0243 | 10.6 | 14.5 | 35.3 | 5.40 | 0.952 | 0.0406 | 117 | -0.0339 | 0.253 | 0.124 | 2076 | 9.8 | | | 0 | 4352 | 4156 | 19.0 | 2496 | 256 |
| | 2 | 2.28 | 0.0241 | 10.6 | 14.6 | 35.3 | 5.41 | 0.952 | 0.0406 | 117 | -0.0339 | 0.253 | 0.125 | 2078 | 9.8 | | | 0 | 4820 | 4616 | 19.2 | 2551 | 264 |
| | 1 | 2.25 | 0.0222 | 10.5 | 15.0 | 35.2 | 5.49 | 0.952 | 0.0411 | 118 | -0.0340 | 0.257 | 0.128 | 2100 | 10.13 | 2350 | 10.09 | 0 | 38410 | 2315 | 16.5 | 3490 | 319 |
| | 0 | 2.22 | 0.0214 | 10.5 | 15.2 | 35.2 | 5.60 | 0.952 | 0.0413 | 117 | -0.0339 | 0.259 | 0.130 | | | | | 10.2 | | 2276 | 16.5 | | 331 |

*Scenarios ending in extinction are shaded. Empty cells indicate the absence of this category. See Table 1 for the explanation of model parameters.

**The first maximum of the curve is indicated.

‡ If the year is not specified for the ( $Q - Q_w$ ) maximum, then this is a plateau (formally year = ∞).

## 6.3. Universality

Population maximums and minimums are functions of the parameters $\nu$ and $\beta$. However, the calculations reveal universality, which consists in the fact that the ratios $N_{max}(\nu,\beta)/N_{plat}(\nu,0)$ and $N_{min}(\nu,\beta)/N_{plat}(\nu,0)$ are independent of $\nu$, where $N_{plat}(\nu,0)$ is the plateau level for $\beta = 0$ when the population is insensitive to temperature rise (the plateau is understood as the limit of $N$ at $t \to \infty$). The same universality applies to the extremum year (year$_{max}$ and year$_{min}$), which also depends on $\beta$ but does not depend on $\nu$. The law of universality is illustrated in Fig. 15. Corresponding regressions and determination coefficients are indicated on the panels. An increase in the loss coefficient $\nu$ is compensated by an increase in the plateau level: a larger population contributes to a decrease in total losses since $N$ is in the denominator of the corresponding term in equation (23) (or $z$ in the normalized equation (30)). The graphs show that with a decrease in sensitivity in the limit $\beta \to 0$, the maximum and minimum converge in 2214 and annihilate.

## 6.4. Population extremums along the critical curve

Population extremums and the years of their appearance change when moving along the critical curve as shown in Fig. 16. For a low loss coefficient $\nu$, the extremums $N_{max}$ and $N_{min}$ differ significantly. For $\nu = 0.01$, minimum is present only on the subcritical curve (in Fig. 12 this is the curve $\beta = 3.39$, while the supercritical curve $\beta = 3.40$ close to it does not have minimum). However, with increasing $\nu$ (but in the area of $\nu < \nu*$ where the critical curve exists), minimums appear on both adjacent curves.

For $\nu = 0.04$, the corresponding example is shown in Fig. 13: these are the curves $\beta = 2.96$ and 3. Further, the maximum and minimum gradually approach each other and coincide at the endpoint $\nu*$, reaching the value $N = 10.2$ billion. For higher values of $\nu$, there are no extremums; only the scenario $\beta = 0$ remains sustainable and reaches a plateau, which rises with $\nu$ as shown in Fig. 15. The moments of the appearance of extremums for small $\nu$ are quite far from each other (500 years apart for $\nu = 0.01$, see Fig. 16, lower panel). With increasing $\nu$, they gradually get closer and finally coincide (year$_{max}$ = year$_{min}$ = 2214) at the endpoint of the critical curve.

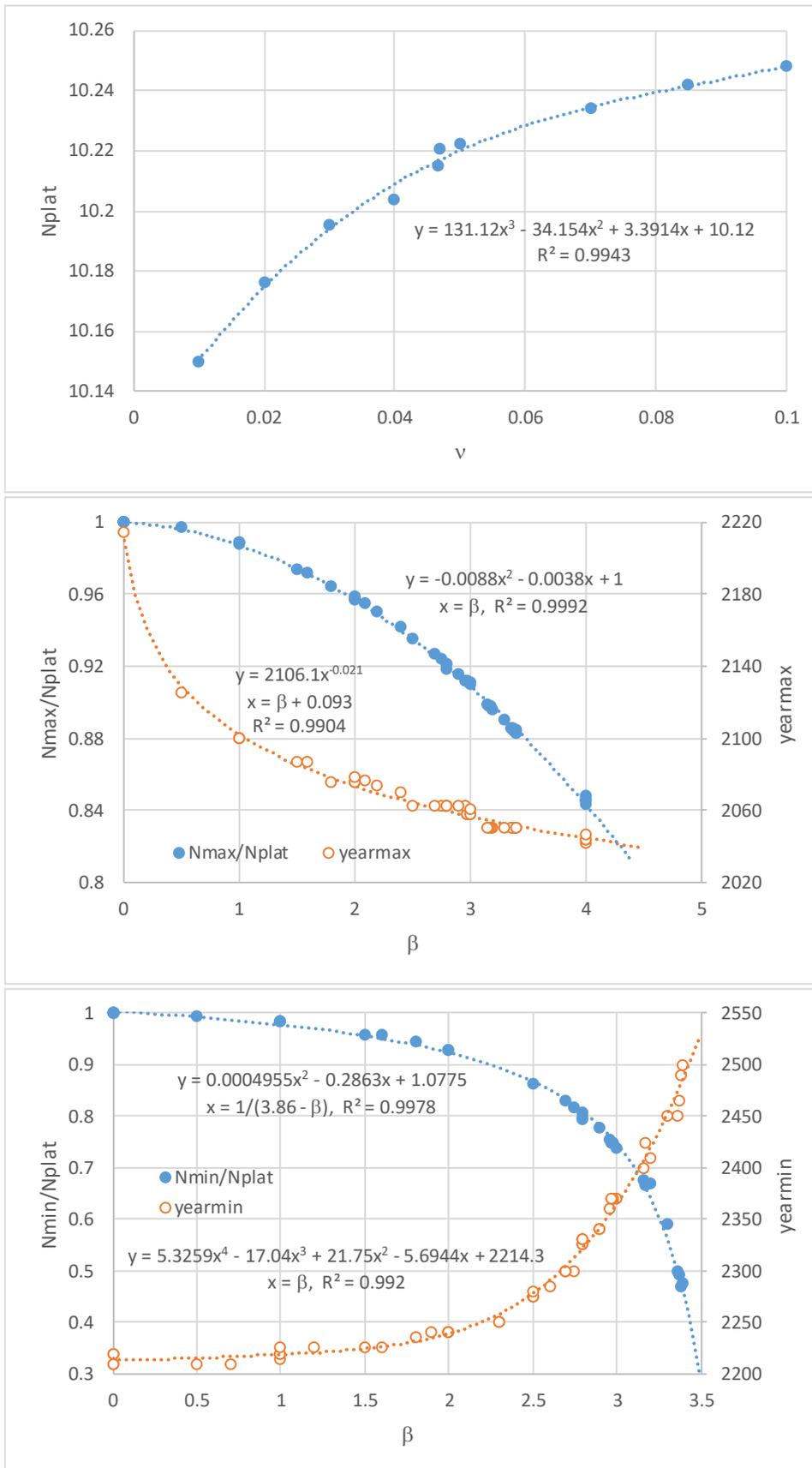

Fig. 15. Stationary population $N_{\text{plat}}$ for $\beta = 0$ as a function of $\nu$ (top panel); population maximum and minimum and years of their appearance as a function of $\beta$ (middle and lower panels).

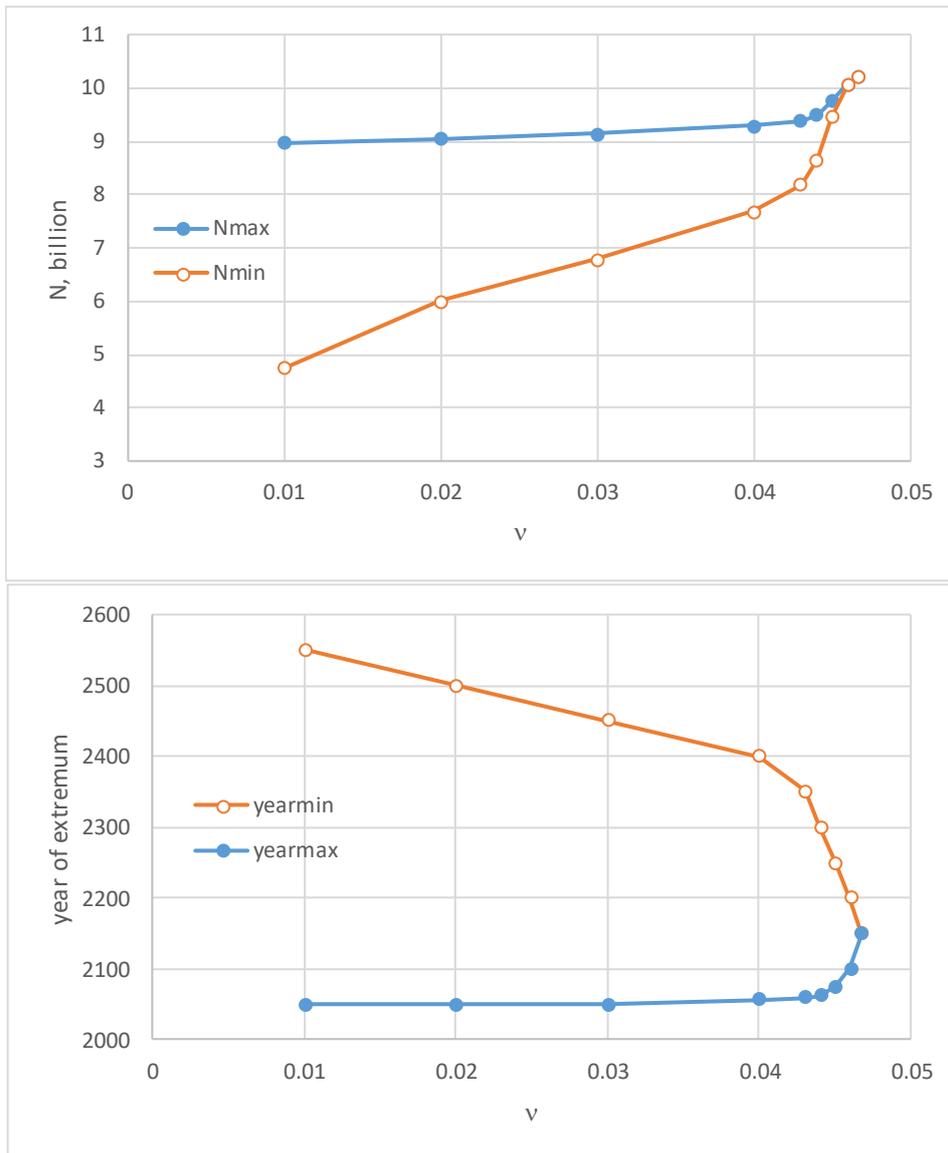

Fig. 16. The first population maximum and minimum and the year of their appearance when moving along the critical curve $\beta(\nu)$.

## 6.5. Parameters along the critical curve

The model parameters change when moving along the critical curve as shown in Fig. 17. It is seen that the curves experience a kink at the endpoint $\nu*$ of the critical curve. In some cases, the kink is negligible, such as for $Q_s$, $u_{in}$ and $\alpha$. The presence of a kink is associated with the loss of stability and the disappearance of civilization with finite temperature sensitivity $\beta > 0$. With a high loss coefficient $\nu \geq \nu*$, survival is only possible with absolute insensitivity to temperature rise ($\beta = 0$).

For population $N$, the relative error in the data is estimated as $4-5\%$ (Burch, 2015). The accuracy of the other data is probably not better. For $N$, this gives an absolute error of at least 0.3 billion for the current population of 7.7 billion, while calculations deviate from the data to a much lesser extent of about 0.08 billion. Thus, the accuracy of the model calculations is quite acceptable.

## 6.6. Changing the share of fossil fuels

The results of calculating the share $f$ of fossil fuels in total energy consumption are shown in Fig. 18. The scenario groups for different values of the loss coefficient $\nu$ contain subcritical and supercritical scenarios. The first category includes: $\beta=0$ and 2.8 for $\nu=0.01$; $\beta=0$ and 2.7 for $\nu=0.04$; and only $\beta=0$ for $\nu=0.1$. The remaining scenarios fall into the second category. In subcritical scenarios, the share of fossil fuels decreases, asymptotically tending to zero. In supercritical scenarios, the decline in $f$ slows down, and at some point, growth begins, which ultimately leads $f$ to the initial pre-industrial level. This behavior is caused by a loss of knowledge, which forces the remaining population to use readily available fuels, the production of which does not require complex technologies. In the later stages of the process, the loss of knowledge is accompanied by a reduction in the population to complete extinction. The minimum value of $f$, the moment it appears, and, for comparison, the moment of extinction are listed below for some scenarios:

| $\nu$ | $\beta$ | $f_{\min}$ | year$_{\min}$ | year$_{\text{ext}}$ |
|-------|---------|------------|---------------|---------------------|
| 0.01  | 4       | 0.348      | 2182          | 2348                |
|       | 3.4     | 0.067      | 2457          | 2940                |
| 0.04  | 4       | 0.417      | 2149          | 2378                |
|       | 3       | 0.034      | 3375          | 5127                |
| 0.1   | 4       | 0.537      | 2111          | 2349                |
|       | 3       | 0.315      | 2197          | 2689                |
|       | 2       | 0.089      | 2551          | 4352                |

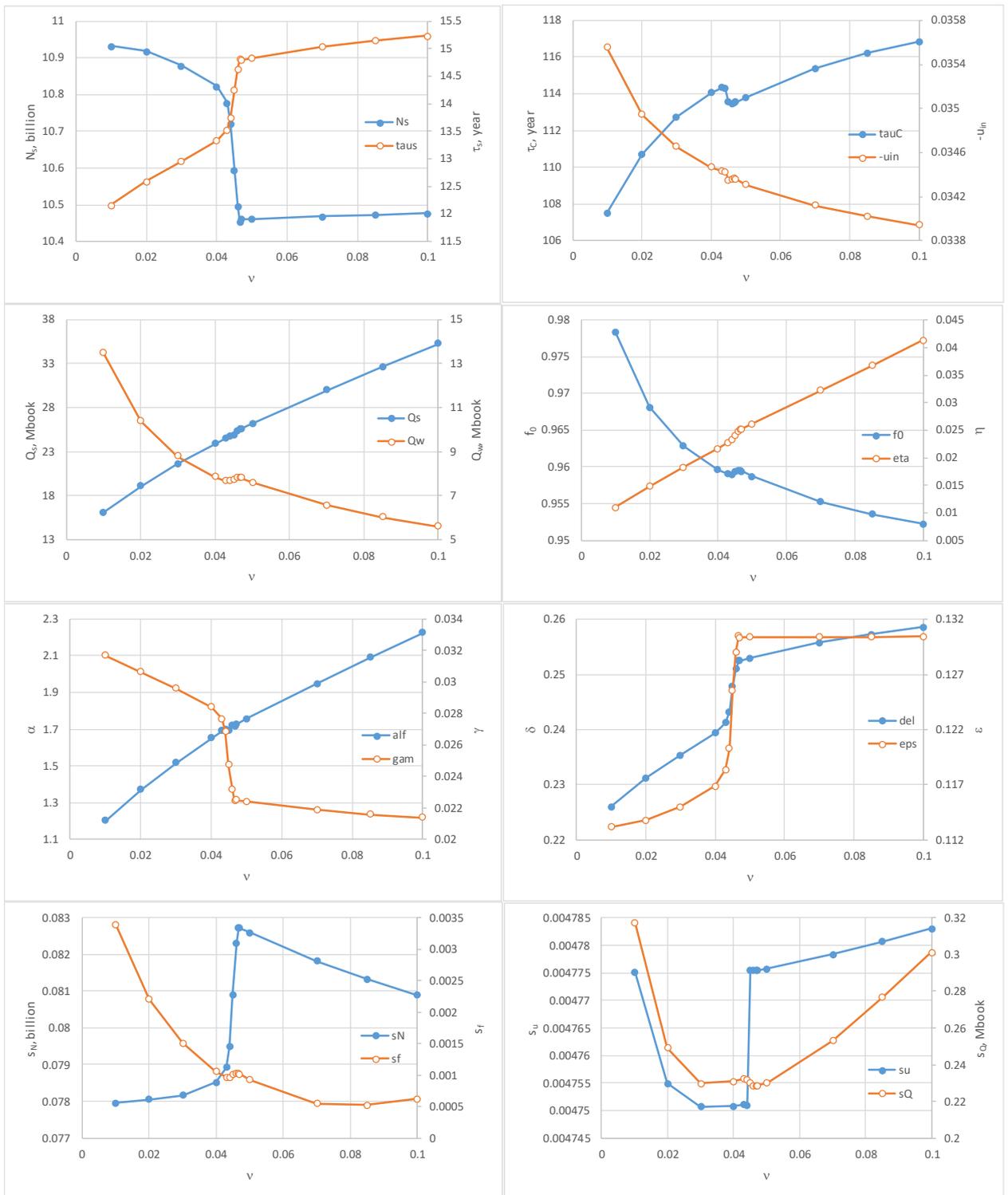

Fig. 17. Model parameters and standard deviations when moving along the critical curve β(ν) . See Table 1 for notations.

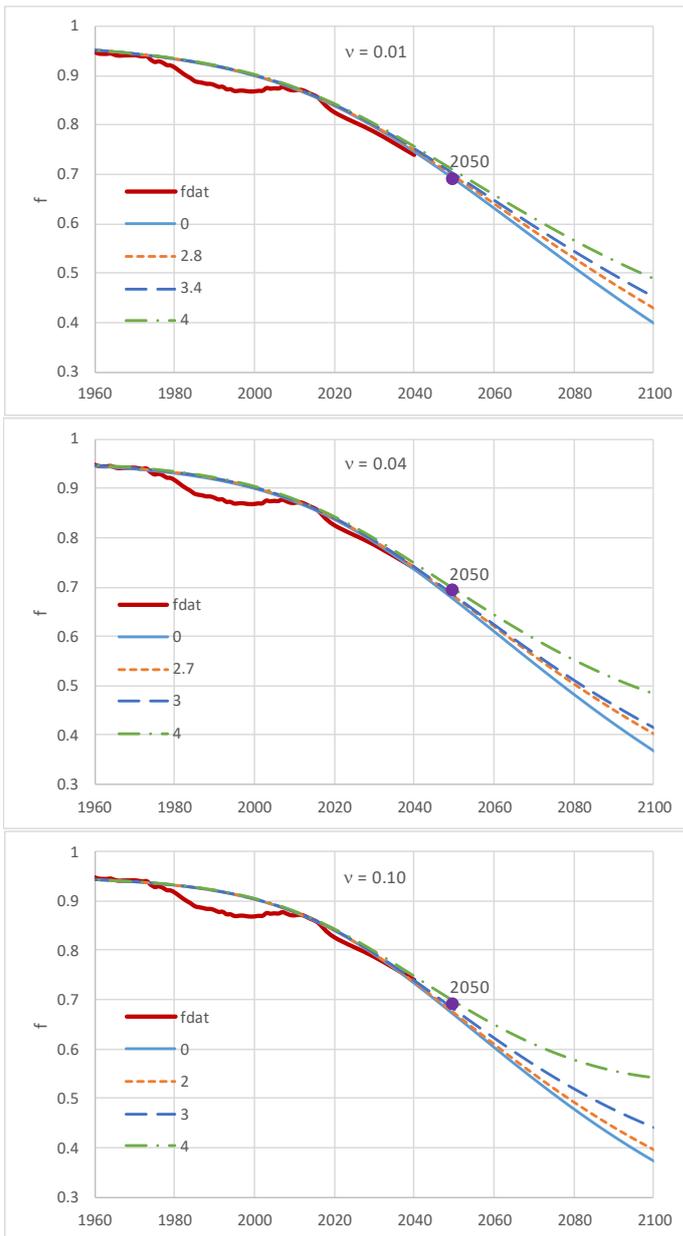

Fig. 18. The share of fossil fuels in total energy consumption for different values of ν and β according to the results of model calculations. For comparison, BP data and BP projection up to 2040, as well as our projection for 2050 (see Section 5.2) are shown. Legend: data and values of β .

## 7. Discussion

### 7.1. Physical model parameters

We begin the discussion with the behavior of physical (dimensional) parameters. Given the scale factors (29) and dimensionless parameters (34) calculated earlier for various scenarios, we

can write

$$\sigma = \frac{\vartheta}{\beta}, \;\; w = \frac{Q_s}{N_s \tau_s}, \;\; v = \frac{1}{b_s} = \frac{Q_s}{\tau_s}, \;\; m_s = \frac{1}{h} = \frac{\nu N_s}{\tau_s},$$

$$\tau_d = \frac{1}{l_s} = \frac{\tau_s}{\gamma}, \;\; Q_a = \frac{1}{a} = \frac{Q_s}{\alpha}, \;\; Q_c = \frac{1}{\sqrt{c}} = \frac{Q_s}{\sqrt{\eta}}, \tag{45}$$

where $\sigma$ is the temperature tolerance, $w$ is the specific rate of knowledge production. New parameters that are easier to interpret have also been introduced: $v$ is the scale of knowledge production rate, $m_s$ is the scale of mortality associated with the loss of knowledge, $\tau_d$ is the characteristic time of energy dissipation in knowledge production, $Q_a$ is the knowledge amount, upon reaching which the inhibition of knowledge production due to environmental limitations becomes noticeable (regarding inhibition see Sections 2.1 and 2.2), $Q_c$ is the knowledge amount that is needed to reduce the share of fossil fuels in total energy consumption through the transition to alternative (non-hydrocarbon) energy sources. The change in these parameters when moving along the critical curve is shown in Fig. 19.

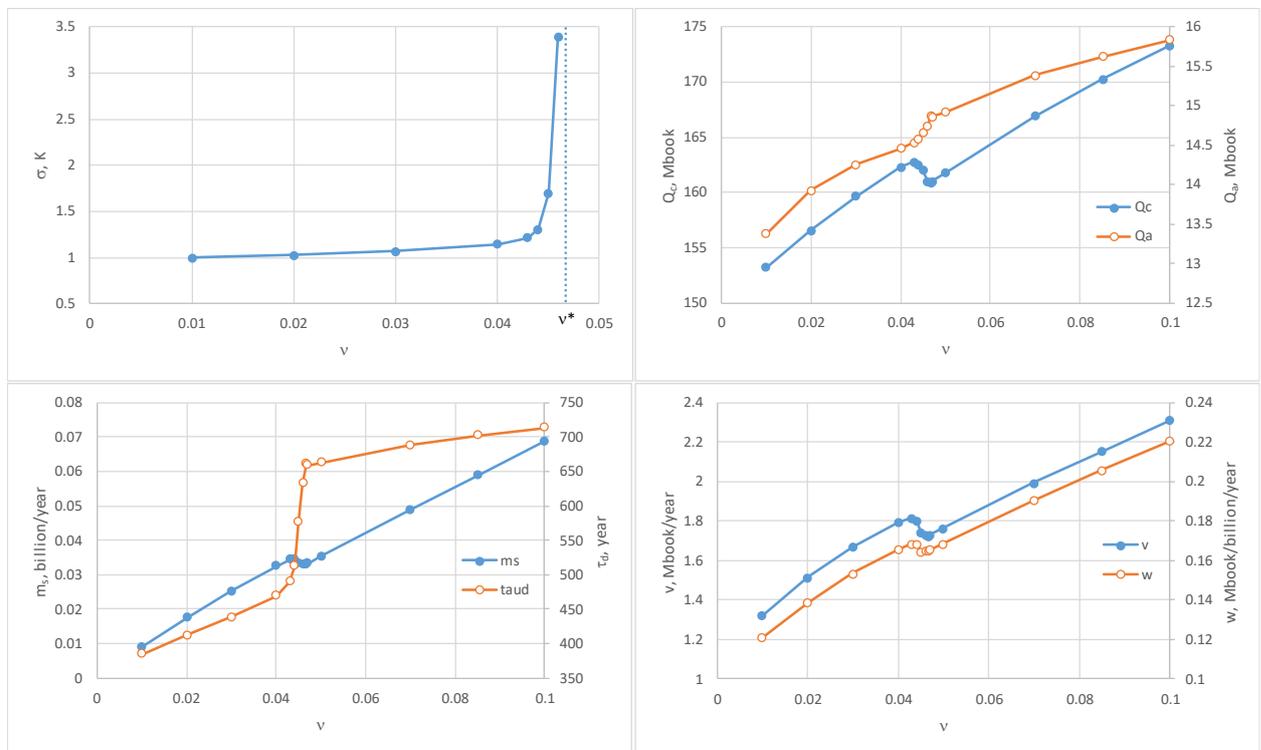

Fig. 19. Physical parameters along the critical curve.

The temperature tolerance $\sigma$ is close to 1°C far from the endpoint $\nu* = 0.0467$, but $\sigma$ increases unlimitedly as $\nu$ approaches $\nu*$, where a sustainable civilization must be insensitive to temperature rise.

The scales of knowledge $Q_a$ and $Q_c$ should increase with increasing the loss coefficient $\nu$, which is confirmed by calculations (except for the neighborhood of $\nu*$). Moreover, it turns out that $Q_c$ is an order of magnitude larger than $Q_a$. This can be explained as follows. Civilization in its development and accumulation of knowledge disturbs the environment, and when the negative feedback becomes significant (at $Q \sim Q_a$), civilization is forced to produce a huge amount of knowledge ($Q \sim Q_c$) in order to create new environment-friendly technologies and restore the environment.

With an increase in the loss coefficient $\nu$, the mortality rate $m_s$ increases almost linearly from 0.01 to 0.07 billion/year in the considered range of $\nu$; linearity is violated in a small neighborhood of the point $\nu*$, where a kink of the curve is observed.

The characteristic energy dissipation time $\tau_d$ varies significantly when passing through the point $\nu*$: from 390 – 470 years for $\nu \leq 0.04$ to 660 – 720 years for $\nu > \nu*$.

The knowledge production rate $v$ and the specific rate of this process $w = v/N_s$ increase with increasing the loss coefficient (except for the vicinity of the point $\nu*$), and both curves are almost parallel since $N_s$ changes insignificantly, only by 5% (see Fig. 17). According to Fig. 19, the scale $v$ of knowledge production rate varies in the range 1.3 – 2.3 Mbook/year depending on the loss coefficient: larger losses require more production of knowledge when moving along the critical curve (recall that the unit "book" corresponds to the books selected for the book stock of the Library of Congress, see Section 5.4). Based on 1 billion people, 0.12 to 0.22 Mbook/year is produced (see curve $w$).

### 7.2. Civilization at the edge of stability

As shown in Sections 5 and 6, two control parameters – coefficient of knowledge loss $\nu$ and sensitivity of population to temperature rise $\beta$ – determine all other parameters and thereby control the system dynamics. Let us estimate the most probable values of $\nu$ and $\beta$. To do this, we introduce the total deviation

$$S_{\text{total}} = \frac{1}{4} \sum_X \frac{s_X}{X_{\text{aver}}}, \quad X = N, f, u, Q, \tag{46}$$

where $s_X$ and $X_{\text{aver}}$ are the standard deviation and the average value of $X$:

$N_{\text{aver}} = 4.85$ billion, $f_{\text{aver}} = 0.907$, $u_{\text{aver}} = 0.262$ (each over $1950 - 2017$),

$Q_{\text{aver}} = 7.78$ Mbook (over $1898 - 1982$);

standard deviations in physical units are shown in Fig. 17 (bottom panels). The total deviation is presented in Fig. 20.

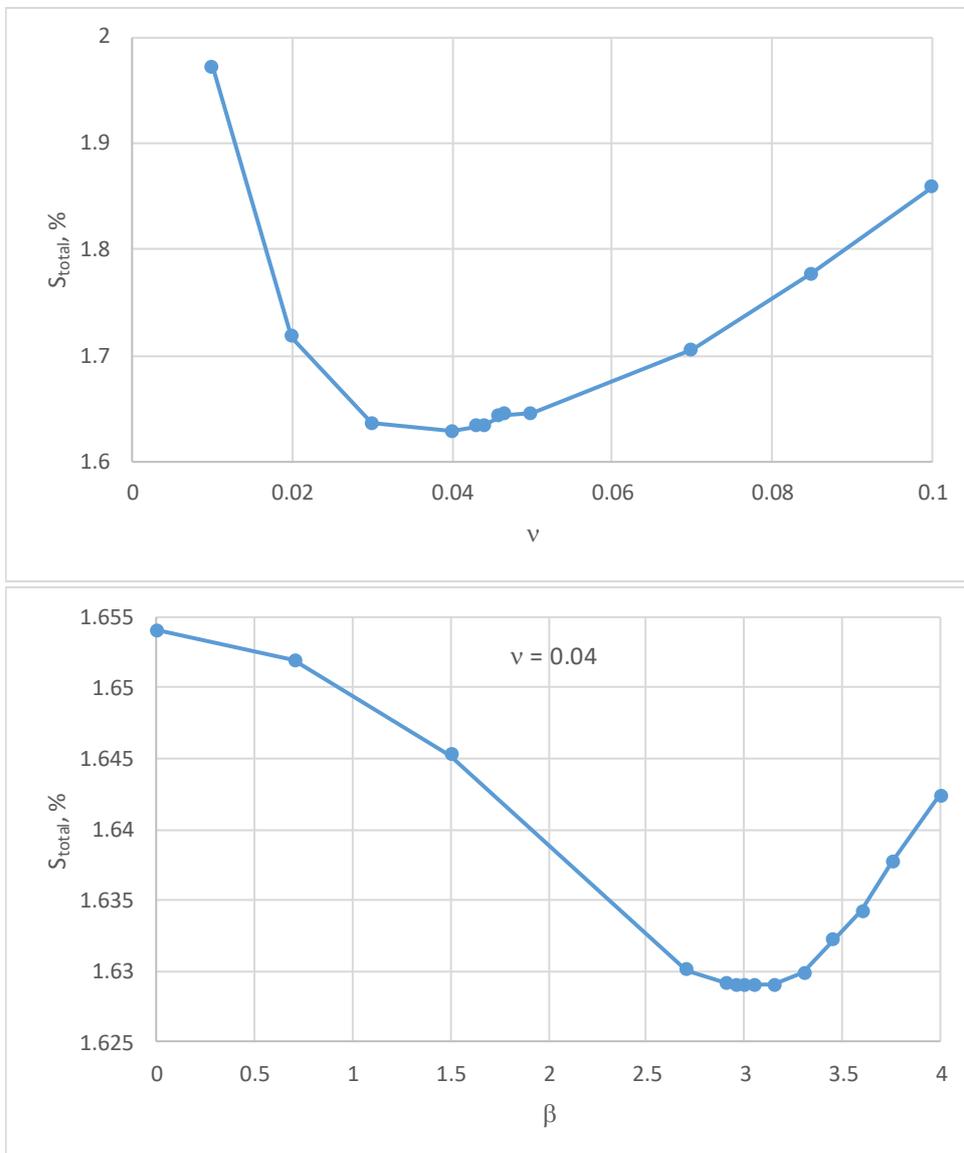

Fig. 20. Total deviation along the critical curve (upper panel) and along the cross-section $\nu = 0.04$ (lower panel).

$S_{\text{total}}$ changes along the critical curve as shown in the upper panel; the minimum is reached at $v_c = 0.04$. The change in $S_{\text{total}}$ across the critical curve in the cross-section $v = 0.04$ is shown in the lower panel; the minimum is reached at $\beta_c = 2.96+$ (between 2.96 and 2.97). Thus, the optimum point

$$(v_c, \beta_c) = (0.04,\ 2.96+), \tag{47}$$

as the most likely place for our civilization, lies right on the critical curve (see Fig. 10). This situation is unstable: a small decrease in $\beta$ transfers civilization to the stable area where the population eventually reaches a plateau of 10.2 – 10.4 billion. On the contrary, with a small increase in $\beta$, civilization falls into the unstable area where it disappears over time. Its lifetime depends on the depth of invasion of the unstable area: the farther from the critical curve, the shorter the lifetime (Table 3). So, with $\beta = 2.97$, civilization disappears in 20292, and with $\beta = 3$ much earlier, already in 5127. For the scenarios $\beta = 2.96$ and 2.97, population maximums coincide: 9.3 billion in 2058; minimums are also the same, 7.7 billion, but the years of their appearance are slightly different: 2381 and 2391. The first temperature maximums are the same, 16.4°C; they fall on 2320 and 2341. In the supercritical scenario $\beta = 2.97$, there is a second temperature maximum of 17.1°C, which falls on 20181; and 111 years after that, civilization disappears. In the short term (by the year 2100), both scenarios give a population of 8.9 billion and a temperature of 16.0°C (against the current 15°C).

Let us consider a group of scenarios in the range $v = 0.03 – 0.044$ and $\beta = 2.7 – 3.3$ around the optimum point $(v_c, \beta_c)$ corresponding to the level $S_{\text{total}} = 1.63\%$ (Fig. 20). They give the following ranges of the population: 9.1 – 9.4 billion in 2050, and 8.5 – 9.3 billion in 2100. The population maximum of 9.1 – 9.5 billion appears in the interval 2054 – 2062. All scenarios from this group give the same temperature in the corresponding year: 15.4°C in 2050 and 16.0°C in 2100. In other scenarios, the variables change as shown in Fig. 13.

Note that there are no intermediate stable states: the population either reaches the upper plateau or vanishes. So, we can conclude that civilization is at the edge of stability.

Relations (45) and the obtained values of $v$ and $\beta$ allow us to find the temperature tolerance and the scale of mortality responsible for the loss of knowledge: $\sigma = 1.14$ K and $m_s = 0.032$ billion/year. If instead of the population scale $N_s$ in the definition of $m_s$ we use the current population $N = 7.7$ billion, then we get $m = 0.046$ billion/year.

Let us compare $m$ with real mortality. It is known (UN, 2019) that crude death rate is now 7.6 per 1000 per year that yields mortality $m_0 = 0.059$ billion/year. The inequality $m < m_0$ means that knowledge is lost more slowly than knowledge holders die, which can be explained by the transfer of knowledge to the next generations through learning. Otherwise, when $m > m_0$, knowledge is lost faster than holders die; this may be, for example, due to problems with education.

The wider and better the education, the smaller the $m$ (and also the $m_s$). Accordingly, the loss coefficient ν also decreases, so that with the same temperature sensitivity β, civilization deepens into the stable area (Fig. 10), which confirms the well-known truth: improving education increases the stability of civilization. On the contrary, with degrading education, the coefficient ν increases, and civilization shifts to the area of instability. Thus, the control parameter ν reflects the capabilities of the education system as a knowledge dissemination tool. This opens up the possibility of expanding our model in two ways: (i) by including in the model a dynamic equation for ν which describes processes in the education system; or (ii) by introducing an objective functional depending on ν, and then solving the optimization problem. Meanwhile, in this study, the parameter ν is constant throughout a scenario but changes when moving from one scenario to another.

## 8. Conclusion

A model of the civilization–environment system has been developed, in which the dynamics of the following macro-variables are tracked: world population, the amount of knowledge used, the share of fossil fuels in world energy consumption, $CO_2$ concentration in the atmosphere, and global mean surface temperature. Distinctive features of this model are as follows. The equation for knowledge production rate takes into account energy dissipation and its relationship with $CO_2$ concentration and temperature. The concept of active knowledge circulating in society and determining the level of life-support technologies is introduced in the model. Knowledge is produced by humanity (which acts as a combination of producers and holders of knowledge) at a rate proportional to the population. At the same time, knowledge is lost due to mortality at a rate inverse to the population. The $CO_2$ balance equation considers emissions from burning fossil fuels and also reverse processes — land uptake and ocean invasion. The dependence of $CO_2$ emissions on the population and the amount of active knowledge has been established.

The model developed was calibrated using literature data for all calculated variables. In calibration, two control parameters remained free: knowledge loss coefficient $\nu$ and sensitivity of population to temperature rise $\beta$, which determine all other model parameters. Scenarios corresponding to different values of the control parameters well consistent with historical data but give different trends in the future. In total, about 90 scenarios were calculated.

It was found that in space $(\nu, \beta)$ there are two areas separated by the critical curve $\beta(\nu)$: one corresponds to sustainable development, and the other to loss of stability. Calculations show that the total deviation of the model from empirical data reaches a minimum for the scenario which corresponds to the point $(\nu, \beta) = (0.04, 2.96+)$ located right on the critical curve. This point determines the most probable state of our civilization. Location on the critical curve is unstable: a small deviation can eventually lead to either stabilization of the population at the level of 10+ billion, or the complete extinction of civilization. There are no intermediate steady states.

Analysis of various scenarios shows that until 2100 it is already difficult to change the dynamics of temperature rise due to the inertia of the world economy since the share of fossil fuels in energy consumption decreases very slowly: 85% at present, ~70% by the mid-century, and ~40% by the end-century.

The model shows the danger of a high loss of knowledge since this affects the development of environment-friendly life-support technologies and can lead to the extinction of civilization. Sustainable development is possible only at a sufficiently low knowledge loss, which can be achieved by improving public education.

It is worth emphasizing that the main result of this study is not so much a quantitative prediction of the future, the accuracy of which is unclear because of the unavoidable sketchiness of the model, as a qualitative picture revealing the interconnection of different factors and showing trends in the development of civilization. An important conclusion is that civilization is at the edge of stability.

The approach proposed may also be useful as part of the scenario planning technique for the future, the main idea of which was expressed by Robert Costanza:
"Predicting the future is impossible. But what we can do is layout a series of plausible scenarios,

which help to better understand future possibilities and the uncertainties surrounding them."
(Costanza, 2013).

**Conflicts of interest**: The author declares no conflict of interest.

**About the author:**

Boris M. Dolgonosov, Ph.D. (1979, Moscow State University), D.Sc. (1997, Russian Academy of Sciences), head of the Ecological modeling laboratory, Institute of Water Problems, Russian Academy of Sciences, Moscow — until September 2014. Currently: Independent researcher, Haifa, Israel; the author of more than 160 scientific publications and two monographs.